\documentclass[twocolumn,prl,aps,superscriptaddress,10pt]{revtex4-1}
\usepackage{amsmath}
\usepackage{amssymb}
\usepackage{amsfonts}
\usepackage{bm}
\usepackage{amssymb}
\usepackage{graphicx}
\usepackage{dcolumn}
\usepackage{txfonts}
\usepackage{makeidx}
\usepackage{color}
\usepackage{mathtools}
\usepackage{threeparttable}
\usepackage{soul}
\usepackage{hyperref}
\hypersetup{colorlinks = true, allcolors = blue}
\bibpunct{\color{blue}[}{\color{blue}]}{,}{n}{}{;}
\usepackage{breakurl}
\usepackage{cleveref}
\usepackage[all]{hypcap}

\begin{document}

\title{Magnetic Noise Enabled Biocompass}
\author{Da-Wu Xiao}
\author{Wen-Hui Hu}
\affiliation{ Beijing Computational Science Research Center, Beijing 100193, China}%
\author{Yunfeng Cai}
\affiliation{ Cognitive Computing Lab, Baidu Research, Beijing 100085, China}
\author{Nan Zhao}
\email{nzhao@csrc.ac.cn}
\affiliation{ Beijing Computational Science Research Center, Beijing 100193, China}%

\date{\today}

\begin{abstract}
The discovery of magnetic protein provides a new understanding of a biocompass  at the molecular level.
However, the mechanism by which magnetic protein enables a biocompass is still under debate, mainly because of the absence of permanent magnetism in the magnetic protein at room temperature.
Here, based on a widely accepted radical pair model of a biocompass, we propose a microscopic mechanism that allows the biocompass to operate without a finite magnetization of the magnetic protein in a biological environment.
With the structure of the magnetic protein, we show that the magnetic fluctuation, rather than the permanent magnetism, of the magnetic protein can
enable geomagnetic field sensing.
An analysis of the quantum dynamics of our microscopic model reveals the necessary conditions for optimal sensitivity.
Our work clarifies the mechanism by which magnetic protein enables a biocompass.
\end{abstract}

\maketitle

\emph{Introduction.}---
Experiments have shown that migrating birds employ the geomagnetic field for orientation and navigation \cite{Phillips_1992,Wiltschko_2005}.
To understand the physical origin of the navigation of animals, several physical models \cite{Johnsen_2005,Wiltschko_2006} have been proposed.
A widely accepted model is the radical pair model, suggested by Ritz \emph{et al.} in Ref. \cite{Ritz_2000}.
This model assumes that the navigation process is governed by radical pairs,
with each pair consisting of an unpaired electron spin \cite{Schulten_1978}. The pairs are usually created via photon excitation and form a spin singlet state \cite{Wiltschko_1993,Engel_2007,Ritz_2004}.
In the geomagnetic field and the magnetic field provided by the local molecular environment, the spin singlet state undergoes a transition to spin triplet states \cite{Ritz_2004,Hiscock_2017}.
The radical pair is metastable and eventually produces
different chemical products according to the spin states of the radical pair \cite{Hore_2016,Maeda_2008,Cai_2010}, and the chemical products determine the subsequent navigation behavior \cite{Ritz_2000,Hore_2016}.

In the radical pair model, we focus on the singlet-triplet interconversion mechanism at the molecular level.
Homogeneous geomagnetic fields cannot change the spin singlet/triplet state because of the conservation of the total spin angular momentum.
Only inhomogeneous magnetic fields can cause transitions between the spin singlet and the spin triplet states. Microscopically, inhomogeneous magnetic fields are provided by the surrounding magnetic moments (either nuclear spins or electron spins) in biological molecules.
Through the interaction between these spins, the radical pair can feel an effective magnetic field.
Nevertheless, the detailed microscopic origin of the singlet-triplet interconversion process remains unclear.

Previous studies have focused on the nuclear spin environment around the radical pair.
For example, experiments found that the ${\rm FADH^{\bullet}-O_{2}^{\bullet-}}$ molecule (which couples the radical pair via the hyperfine interaction) is relevant to animal navigation \cite{Ritz_2009,Mouritsen_2004,Liedvogel_2007,Maeda_2012}.
Theoretically, studies in Refs. \cite{Cai_2010,Cai_2013,Cai_2012,Walters_2014,Solov_yov_2007} showed that the nuclear spin environment is capable of providing local magnetic fields and enabling a biocompass.
Our analysis shows that the nuclear spin concentration and the anisotropic dipolar coupling between the radical pair spins and the bath nuclear spins play important roles to enable a biocompass \cite{Sup}.
In addition to progress on the nuclear spin bath, Ref. \cite{Qin_2015} reported a new putative magnetic receptor (MagR) and showed that the MagR forms a rod-like magnetosensor complex with the radical pair in the photoreceptive cryptochromes.
The MagR consists of an Fe-S cluster protein, with the d electrons in the Fe atom contributing the electron spins \cite{Noodleman_1995,Munck_1997}.
It is reasonable to assume that the navigation behavior arises from the effect of an electronic spin bath.

However, the microscopic role of the magnetic protein is under debate.
In Ref. \cite{Meister_2016}, the author pointed out that electron spins are hardly polarized at room temperature and cannot produce a significant single-triplet transition process.
Therefore, it is crucial to elucidate what makes the biocompass possible in the absence of a finite magnetization in the MagR.
In this paper, we propose that the magnetic field fluctuation, rather than the mean magnetization, is capable of producing the spin singlet and triplet transition.
The electron spin bath of the MagR introduces a fluctuating local magnetic field to the nearby radical pairs via the magnetic dipole-dipole interaction, and this local magnetic field actually enables singlet-triplet interconversion.

First, with a semiquantitative analysis of a radical pair coupling to an electron spin bath in the geomagnetic field,
we find two necessary intuitive requirements for the local magnetic field that need to be satisfied:
i) The strength of the noise magnetic field must be comparable to the geomagnetic field
($\sim10^{-1}\,\rm{Gauss}$), and ii) The local magnetic field should have directional dependence.
Then, we establish a microscopic model that describes the spin dynamics of the radical pair in an electron spin bath.
With theoretical analysis and numerical calculations, we find that the singlet fidelity of the radical pair can exhibit a sensitive geomagnetic field direction dependence.
Our work provides new insights into the understanding of the biocompass mechanism.

\emph{Theoretical Model.}---
We consider a radical pair interacting with a spin bath described by the following Hamiltonian:
\begin{equation}
H = H_{\rm rp} + H_{\rm bath} + H_{\rm int},
\label{eq:TotalHamiltonian}
\end{equation}
where $H_{\rm rp}$, $H_{\rm bath}$ and $H_{\rm int}$ are the Hamiltonians of the radical pair, the bath spin and their interaction, respectively.
The radical pair consists of two electron spins $\mathbf{S}_1$ and $\mathbf{S}_2$,
forming the singlet state $\vert S\rangle =\left(\vert \uparrow\downarrow\rangle - \vert \downarrow\uparrow\rangle\right)/\sqrt{2}$
and triplet states $\vert T_{0}\rangle=\left(\vert \uparrow\downarrow\rangle + \vert \downarrow\uparrow\rangle\right)/\sqrt{2}$,
$\vert T_{+}\rangle = \vert\uparrow\uparrow\rangle$, and $\vert T_{-}\rangle = \vert\downarrow\downarrow\rangle$ \cite{Ritz_2000}.
In the singlet-triplet representation, the radical pair Hamiltonian is diagonalized as
\begin{equation}
\label{eq:HamiltonianRP}
  H_{\rm rp} = \sum_{k} \omega_k \left| \phi_k \right> \left< \phi_k \right|,\quad {\rm for}~\vert\phi_k\rangle \in \left\{\vert S\rangle, \vert T_0\rangle, \vert T_+\rangle, \vert T_-\rangle\right\},
\end{equation}
where $\omega_k$ is the energy of the singlet/triplet state $\vert \phi_k\rangle$.
The radical pair is subjected to a magnetic environment consisting of $N$ interacting spins $\{ \mathbf{J}_i\}_{i=1}^N$.
\begin{equation}
\label{eq:HamiltonianBath}
H_{\rm bath} =  \sum_{i=1}^{N} \gamma_i ~ \mathbf{B} \cdot \mathbf{J}_i + \sum_{i>j=1}^{N} \mathbf{J}_i \cdot \mathbb{D}_{ij} \cdot \mathbf{J}_j,
\end{equation}
where $\mathbf{B}$ is the geomagnetic field, $\gamma_i$ is the gyromagnetic ratio of the $i$-th bath spin, and $\mathbb{D}_{ij}$ is the coupling tensor between $\mathbf{J}_i$ and $\mathbf{J}_j$.
The radical pair spins couple to the bath spins through the interaction Hamiltonian
\begin{equation}
\label{eq:HamiltonianInt}
H_{\rm int} = \sum_{k,i} \mathbf{S}_k \cdot \mathbb{A}_{ki} \cdot \mathbf{J}_i \equiv \sum_{k=1,2} \gamma_e \mathbf{S}_k \cdot \mathbf{b}_k,
\end{equation}
where $\gamma_e$ is the electron spin gyromagnetic ratio, $\mathbb{A}_{ki}$ is the coupling tensor and $\mathbf{b}_k$, as seen by the radical pair spin $\mathbf{S}_k$, is the effective magnetic field caused by the bath spins.

Indeed, Eqs.~\crefrangeformat{equation}{#3(#1)--(#2)#4}\crefrange{eq:TotalHamiltonian}{eq:HamiltonianInt} are quite general Hamiltonians describing the interacting spins.
Since the precise electronic structure of the radical pair and the bath spins of the biocompass system is still unclear,
we did not specify the details of the singlet/triplet energies $\omega_k$ and the concrete forms of the coupling tensors $\mathbb{A}_{ki}$ and $\mathbb{D}_{ij}$ in Eqs.~\crefrangeformat{equation}{#3(#1)--(#2)#4}\crefrange{eq:TotalHamiltonian}{eq:HamiltonianInt}.
However, we assume that the random motion of the spins (typically with a time
scale $> \rm{ms} $ \cite{2016}) is not fast enough to average out the spin dynamics of the radical pair (typically $\sim\rm{\mu s}$ \cite{Hore_2016,Maeda_2008,Cai_2010}).
Nevertheless, we will show that we still need some reasonable assumptions for $\omega_k$, $\mathbb{A}_{ki}$ and $\mathbb{D}_{ij}$ based on the known structure of the magnetic protein
to make the coupled system described by Eqs.~\crefrangeformat{equation}{#3(#1)--(#2)#4}\crefrange{eq:TotalHamiltonian}{eq:HamiltonianInt} exhibit strong sensitivity to the geomagnetic field direction.

\emph{Magnetic Fluctuation.}---
We study the quantum dynamics of the radical pair in an unpolarized spin bath.
The radical pair is initially prepared in a singlet state with $\rho_{\rm rp}(0)=\left\vert S \rangle \langle S \right\vert$,
and the bath spins are in a high-temperature mixed state
\begin{equation}
\rho_{\rm bath}(0) =\bigotimes_{i=1}^{N} \frac{\mathbb I_i}{ \mathrm{Tr} [\mathbb I_i] },\label{eq:BathInitial}
\end{equation}
where $\mathbb I_i$ is the identity operator for the $i$-th spin.
Starting from the initial state $\rho(0)=\rho_{\rm rp}(0)\otimes\rho_{\rm bath}(0)$,
the system evolves to $\rho(t)$ driven by the Hamiltonians in Eqs.~\crefrangeformat{equation}{#3(#1)--(#2)#4}\crefrange{eq:TotalHamiltonian}{eq:HamiltonianInt}.
We focus on the singlet state fidelity $P_S(t) = {\rm Tr}\left[\vert S \rangle \langle S \vert \rho(t) \right]$ of the radical pair
and its dependence on the geomagnetic field direction \cite{Ritz_2000}.

\begin{figure}[t]
\includegraphics[scale=0.3]{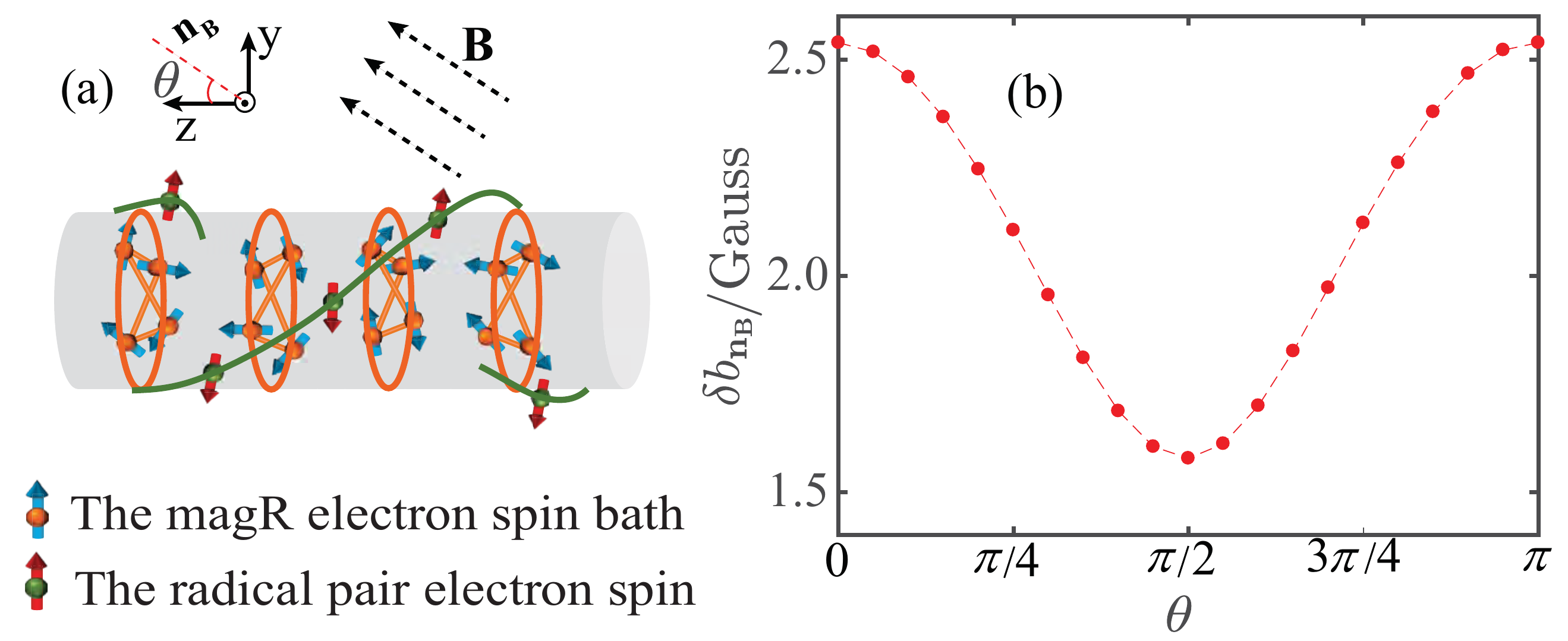}
\caption{\label{fig:figure1}
(color online).
(a) Illustration of the structure of the bath electron spins and the radical pair. The spins are fixed in the proteins.
(b) Magnetic field fluctuation along the geomagnetic field direction $\delta b_{\mathbf{n_B}}$ as a function of the geomagnetic field direction $\theta$.
}
\end{figure}
The field difference $\mathbf{\delta b} =\mathbf{b}_1 - \mathbf{b}_2$ experienced by the two spins of the radical pair causes the singlet-triplet conversion.
Before presenting full quantum mechanical calculations of the singlet fidelity $P_S(t)$, we present a qualitative analysis of the effect of $\mathbf{\delta b}$.
One of the key concerns is that the mean value of the field difference $\mathbf{\delta b}$ vanishes at room temperature, i.e., $\mathrm{Tr} [\rho_{\rm bath}(0)\mathbf{\delta b}]\equiv0$.
This condition strongly challenges the role of the MagR in the biocompass mechanism.
However, the fluctuation of $\mathbf{\delta b}$ can also cause the singlet-triplet conversion.
Specifically, in the following, we consider the variation in the projection of $\mathbf{\delta b}$ along the direction $\mathbf{n}_{\mathbf{B}}$ of the external magnetic field $\mathbf{B}$,
i.e., $\delta b_{\mathbf{n}_\mathbf{B}}^2=\mathrm{Tr} [\rho_{\rm bath}(0)(\mathbf{n_B}\cdot\mathbf{\delta b})^2]$.
Here, $\mathbf{n}_{\mathbf{B}}$ is given by the Euler angle $\theta,\phi$ of $\mathbf{B}$.
We will present requirements for the field difference $\mathbf{\delta b}$ to
play an important role in the biocompass.

First, the fluctuations $\delta b_{\mathbf{n}_B}$ should have comparable strengths to the geomagnetic field $\mathbf{B}$.
In the weak fluctuation limit $\vert \delta b_{\mathbf{n}_B}\vert \ll\vert\mathbf{B}\vert$, the system evolution will be dominated by the homogeneous geomagnetic field $\mathbf{B}$,
and singlet-triplet conversion can hardly occur.
However, in the opposite limit $\vert\delta b_{\mathbf{n}_B}\vert\gg\vert\mathbf{B}\vert$, the geomagnetic field $\mathbf{B}$ will have a negligible influence on the dynamics of $P_S(t)$.
In both limiting cases, the system does not exhibit a biocompass function.
Using the structure obtained in Ref.~\cite{Qin_2015,Sup} and assuming electronic dipolar coupling between the radical pair spins and the bath spins in Eq.~\crefformat{equation}{#2(#1{})#3}\cref{eq:HamiltonianInt},
we find that the magnitude of the coupling tensor $\mathbb{A}_{ki}\sim 10^1~{\rm MHz}$, corresponding to the strength of the fluctuations $\delta b_{\mathbf{n}_B}\sim 10^{-1}~{\rm Gauss}$ [see \crefformat{figure}{Fig.~#2#1{(b)}#3}\cref{fig:figure1}],
is on the same order as the geomagnetic field.

Second, the fluctuation of $\mathbf{\delta b}$ should be sensitive to the direction of the geomagnetic field.
This condition requires the coupling $\mathbb{A}_{ki}$ between the radical pair and spin bath to be anisotropic.
Indeed, the dipolar coupling between the electron spins satisfies this requirement.
Furthermore, the rod-like structure also enhances the anisotropicity of the field fluctuation, since the axial and azimuthal directions are obviously inequivalent.
As an example, \crefformat{figure}{Fig.~#2#1{(b)}#3}\cref{fig:figure1} shows that the fluctuation magnitude changes by a factor of $\sim 2$ as the geomagnetic field direction varies by $\pi$.

With these two intuitive requirements, we find that the dipolar coupling between the radical pair and the MagR spins is a promising candidate to explain the microscopic mechanism of the biocompass.
In the following, we discuss the
optimal conditions of magnetosensation through the quantum
dynamics of the system.

\emph{Optimization of magnetosensation}---
In the system defined by Eqs.~\crefformat{equation}{#2(#1{})#3}\cref{eq:TotalHamiltonian}-\crefformat{equation}{#2(#1{})#3}\cref{eq:HamiltonianInt}, we focus on the dynamics of the singlet fidelity $P_{S}\left(t\right)$ of the radical pair spins.
A full analytical calculation is usually not available for a system of interacting electron spins. Here, we first analyze the short-time behavior of $P_{S}\left(t\right)$.
With the short-time approximation, we obtain the qualitative requirements for $\omega_{k},\mathbb{A}_{ki}$ and $\mathbb{D}_{ij}$ to achieve optimal magnetosensation of the singlet fidelity, which are further confirmed by numerical simulations.

\begin{figure}[b]
\includegraphics[scale=0.19]{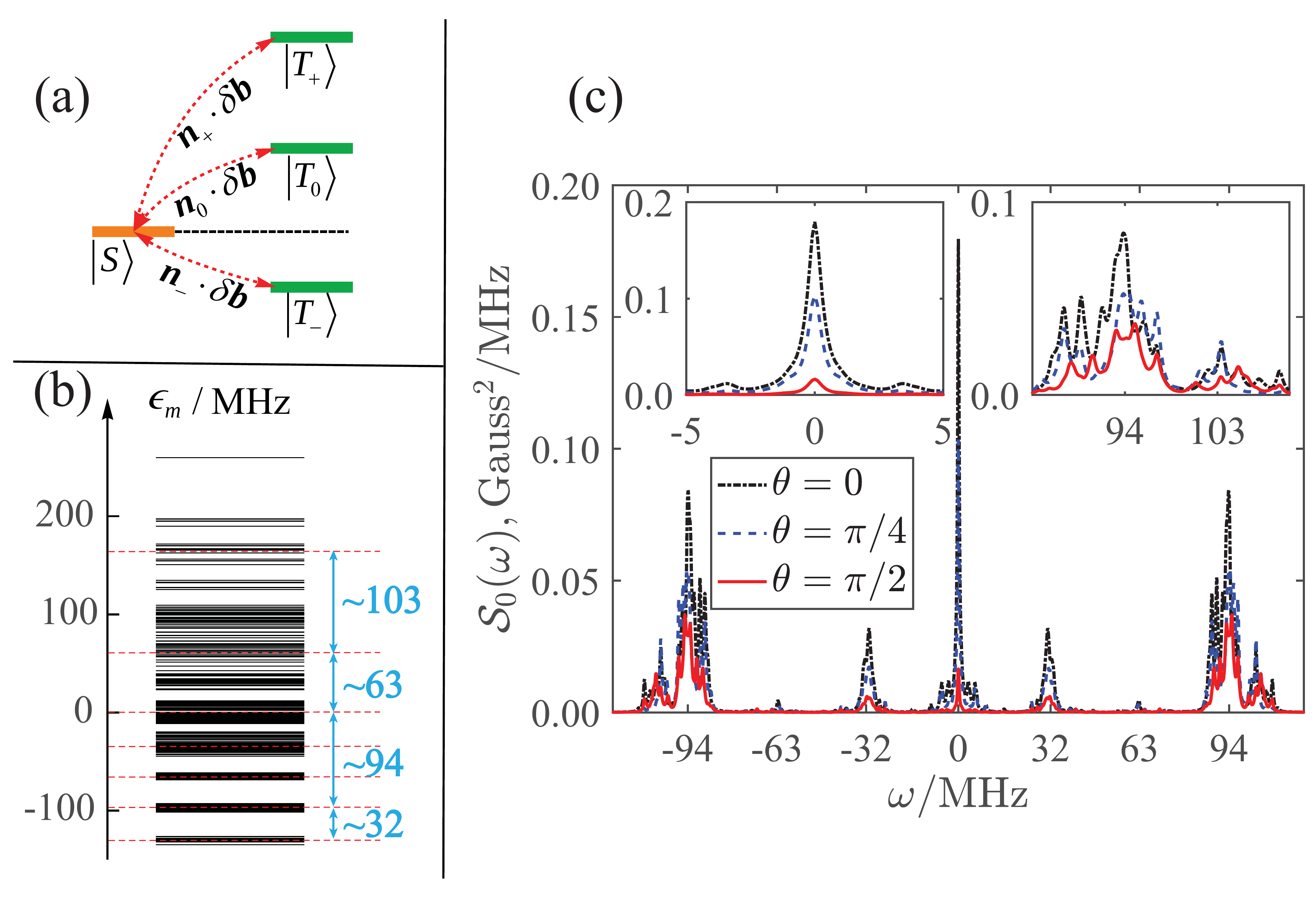}
\caption{\label{fig:figure2} (color online).
(a) Illustration of the radical pair's energy spectrum and the corresponding noise component during the transition.
(b) The energy spectrum of the two rings in the MagR (see text).
(c) The noise spectrum $\mathcal{S}_0(\omega)$ for the two rings in different geomagnetic directions, which shows broadened peaks near $0,32,63,94$, and $\pm103~{\rm MHz}$.
Here, for the convenience of illustration, we add Lorentzian broadening of $0.3~\rm{MHz}\ll\Delta\omega_{peak}$.
}
\end{figure}

The effective field difference $\mathbf{\delta b}$ induces transitions from the singlet state $\vert S \rangle$ to the triplet states $\vert T_{0}\rangle$ and $\vert T_{\pm} \rangle$ and causes a loss of the singlet fidelity.
Specifically, we choose the quantization axis (the $z$ axis) along the direction of the geomagnetic field.
As shown in \crefformat{figure}{Fig.~#2#1{(a)}#3}\cref{fig:figure2}, the longitudinal component $\delta b_z = \delta \mathbf{b} \cdot \mathbf{e}_z$ induces the transition $\vert S \rangle\rightarrow\vert T_0 \rangle$,
while the transverse components $\delta b_{\pm}=\mathbf{\delta b}\cdot\mathbf{n}_{\pm}$ with $\mathbf{n}_{\pm}=(\mp\mathbf{e}_x-i\mathbf{e}_y)/\sqrt{2}$ result in the transitions $\vert S \rangle\rightarrow\vert T_\pm \rangle$. The field difference $\mathbf{\delta b}$ is nonstatic in a full quantum mechanical treatment.
The dynamics of the effective field difference is determined by the interaction within the bath spins as
\begin{equation}
 \mathbf{\delta b}(t)=e^{i H_{\rm{bath}} t}\mathbf{\delta b}(0)e^{-i H_{\rm{bath}} t}.
\end{equation}
The expectation value of the field difference $\langle\mathbf{\delta b}(t)\rangle$ vanishes.
However, $\delta \mathbf{b}(t)$ has finite fluctuations:
$\langle\mathbf{\delta b}(t)\mathbf{\delta b}(t')\rangle={\rm{Tr}} [\mathbf{\delta b} (t)\mathbf{\delta b}(t')\rho_{\rm{bath}}(0)]\neq0$.

To understand the dynamic properties of the field difference $\mathbf{\delta b}(t)$, it is necessary to investigate the interaction within the bath spins.
As illustrated in \crefformat{figure}{Fig.~#2#1{(a)}#3}\cref{fig:figure1}, the MagR spin bath consists of several ring structures.
Within a ring, the distance between the spins is approximately 1-2 nm, while the interring distance is greater than 5 nm (see \cite{Sup} for the coordinates of the electron spins).
Suppose that the spins are all coupled through the magnetic dipole-dipole interaction, i.e., $\mathbb{D}_{ij}=\frac{\mu_{0}\gamma_{e}\gamma_{e}\hbar}{r_{ij}^{3}}\left(1-3\hat{\mathbf{r}}_{ij}\hat{\mathbf{r}}_{ij}\right)$,
with $r_{ij}$ being the distance between two spins and $\hat{\mathbf{r}}_{ij}$ being the unit coordinate vector.
Since the dipolar interaction strength decays as $r_{ij}^{-3}$, the coupling of the spins within a ring ($\sim 10^{1}~{\rm MHz}$) is much stronger than that in different rings ($< 10^0~{\rm MHz}$).
\crefformat{figure}{Figure~#2#1{(b)}#3}\cref{fig:figure2} shows the energy spectrum $\epsilon_m$ of $H_{\rm{bath}}$ obtained by diagonalizing the Schrödinger equation
\begin{equation}
  H_{\rm{bath}}\vert \psi_m\rangle = \epsilon_m \vert \psi_m \rangle,\label{eq:EigenEq}
\end{equation}
with $\vert \psi_m \rangle$ being the eigenstate.
The energy spectrum $\epsilon_m$ forms several discrete bands around $0,\pm32,\pm63,\pm94,$ and $\pm103~{\rm MHz}$, resulting from the strong interaction of the electron spins within a ring.
Each band is further broadened due to the weak interaction of the electron spins between the rings.

With the transition probability $P_{S\rightarrow T_{\alpha}}(t)$ from the singlet state $\vert S\rangle$ to the triplet states $\vert T_{\alpha}\rangle$ (for $\alpha=0$ or $\pm$),
the singlet fidelity is expressed as
\begin{equation}
P_S(t)=1-\sum_{\alpha} P_{S\rightarrow T_{\alpha}}(t).\label{eq:short-time}
\end{equation}
In the short-time limit, the transition probability $P_{S\rightarrow T_{\alpha}}(t)$ is approximated as \cite{Tannoudji_1991,Zhao_2011,Sup}
\begin{equation}
P_{S\rightarrow T_{\alpha}}(t) = \gamma_e^2 \int_{-\infty}^\infty \mathcal{S}_{\alpha}(\omega)  F(t,\omega;\omega_{S T_{\alpha}}) d\omega\label{eq:Approx}
\end{equation}
with $\omega_{S T_{\alpha}}=\omega_S-\omega_{T_{\alpha}}$.
In Eq.~\crefformat{equation}{#2(#1{})#3}\cref{eq:Approx}, the function $\mathcal{S}_{\alpha}(\omega)$ is the power spectrum of the effective field difference $\delta b_{\alpha}$
\begin{equation}
\mathcal{S}_{\alpha}(\omega)=\frac{1}{2^N}\sum_{m,n}\vert \langle\psi_m\vert  \delta b_{\alpha} \vert \psi_n \rangle \vert^2 \delta(\omega-\epsilon_{mn})
\label{eq:NS}
\end{equation}
with $\epsilon_{mn}=\epsilon_m -\epsilon_n$; the function $F(t,\omega;\omega_{S T_{\alpha}})$, defined as
\begin{equation}
F(t,\omega;\omega_{S T_{\alpha}})=\frac{\sin^2\left(\frac{\omega t+\omega_{S T_{\alpha}}t}{2}\right)}{(\omega+\omega_{S T_{\alpha}})^{2}},
\end{equation}
is regarded as a spectrum filter function in the frequency domain, which exhibits a peak centered at $\omega_{S T_{\alpha}}$ with width $\Delta\omega_{\rm{filter}}=1/t$ \cite{Tannoudji_1991}.
Moreover, to directly relate the singlet fidelity $P_S(t)$ to the biochemical process, we define the singlet productivity on a relevant time scale
\begin{equation}
  \Phi_S(\tau;\theta;\phi)= \frac{1}{\tau}\int_0^\tau P_S(t;\theta;\phi) dt,
\end{equation}
which is a function of the geomagnetic direction.
Here, $\tau$ is the relevant time scale in the radical pair model, chosen to be $1~\mu s$ in the subsequent discussion \cite{Hore_2016,Cai_2010}.

The power spectrum $\mathcal{S}_{\alpha}(\omega)$ describes the dynamic property of the field difference $\delta b_{\alpha}$ in the frequency domain.
As an example, \crefformat{figure}{Fig.~#2#1{(c)}#3}\cref{fig:figure2} shows the power spectrum $\mathcal{S}_{0}(\omega)$ [see \cite{Sup} for more results for $\mathcal{S}_{\pm}(\omega)$].
Due to the band structure of the eigenenergies $\epsilon_m$ [see \crefformat{figure}{Fig.~#2#1{(b)}#3}\cref{fig:figure2}], the power spectrum exhibits broadened discrete peaks around specific transition frequencies
[e.g., $\omega_{\rm peak} = 0,\pm32,\pm63,\pm94$, and $\pm103~{\rm MHz}$ for $\mathcal{S}_{0}(\omega)$ shown in \crefformat{figure}{Fig.~#2#1{(c)}#3}\cref{fig:figure2}],
with a typical peak width $\Delta\omega_{\rm{bath}}\sim 10^1 ~ \rm{MHz}$.
Furthermore, due to the anisotropic dipolar coupling between the spins and the rod-like geometric configuration of the MagR, the power spectrum exhibits a dependence on the geomagnetic field direction.
\crefformat{figure}{Figure~#2#1{(c)}#3}\cref{fig:figure2} shows that the amplitudes of the power spectrum peaks of $\mathcal{S}_0(\omega)$ are very sensitive to the different geomagnetic field directions.

\begin{figure}[t]
\includegraphics[scale=0.35]{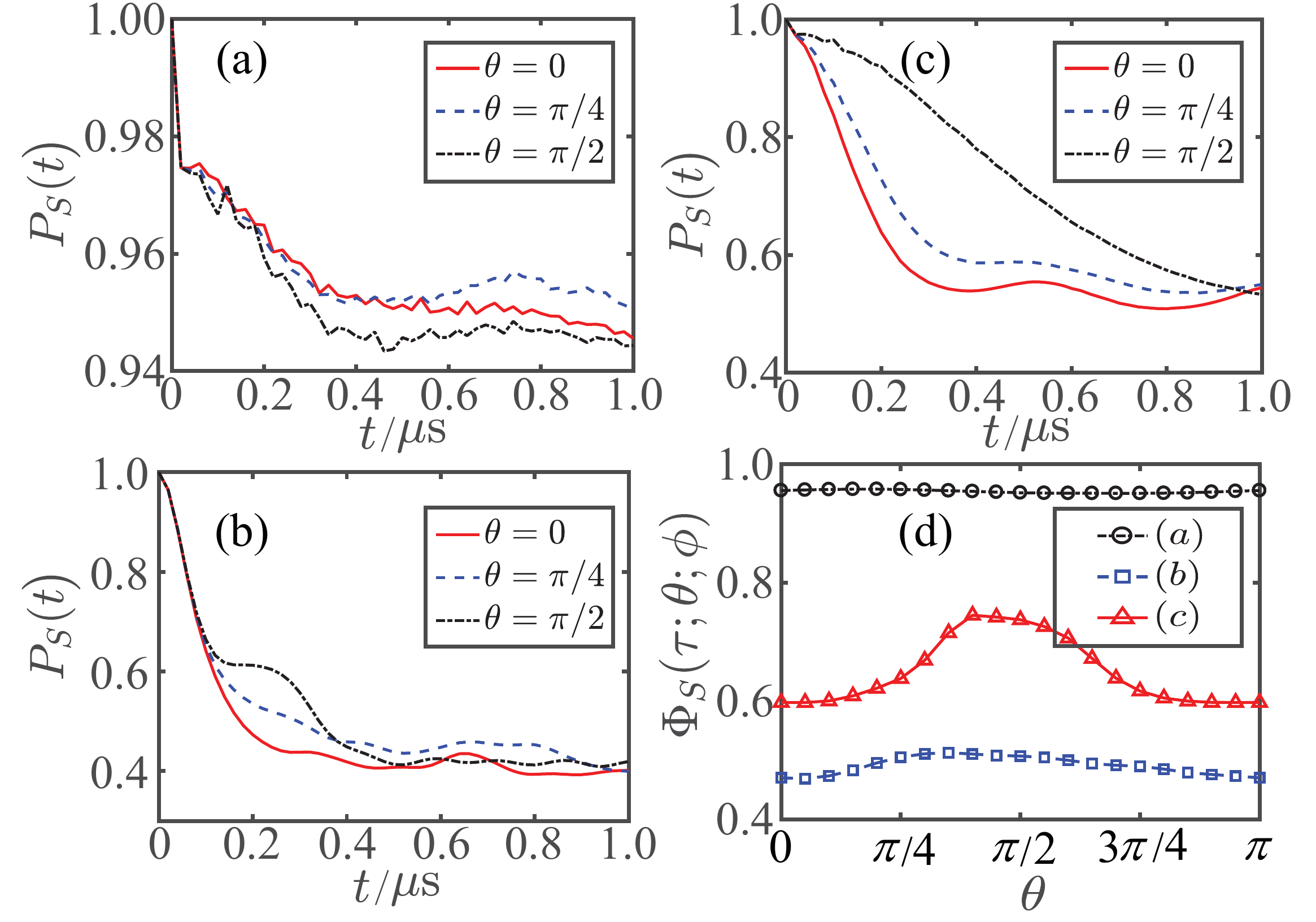}
\caption{\label{fig:figure3} (color online). The numerical result of the singlet fidelity as a function of time for different representative geomagnetic directions:
(a) when all $\omega_{{\rm S T_i}}$ do not overlap with the filter spectrum ($\omega_{{\rm S T_0}}=20~\rm{MHz},\omega_{{\rm S T_+}}=23~\rm{MHz},\omega_{{\rm S T_-}}=17~\rm{MHz}$),
(b) when all $\omega_{{\rm S T_i}}$ overlap with the filter spectrum but all the triplets are degenerate ($\omega_{{\rm S T_0}}=0.1~\rm{MHz},\omega_{{\rm S T_\pm}}=\pm3.0~\rm{MHz}$), and
(c) when $\omega_{{\rm S T_0}}$ overlaps with the noise spectrum $\mathcal{S}_0(\omega)$ while the others do not($\omega_{{\rm S T_0}}=0.01~\rm{MHz},\omega_{{\rm S T_\pm}}=\pm20.0~\rm{MHz}$). (d) The singlet productivity as a function of the geomagnetic field direction $\theta$ for $\phi=0$. \cite{note1}
}
\end{figure}

The overlap between the power spectrum $\mathcal{S}_{\alpha}(\omega)$ and the filter function $F(t,\omega;\omega_{S T_{\alpha}})$ determines the loss of the singlet fidelity, as shown in Eq.~\crefformat{equation}{#2(#1{})#3}\cref{eq:Approx}.
With this observation, we propose the following necessary conditions for a robust biocompass to exhibit a strong dependence on the geomagnetic field direction.

First, at least one of the peaks of the power spectrum must be in resonance with the singlet-triplet transition, i.e.,  $\vert \omega_{S T_i} - \omega_{\rm{peak}} \vert<\Delta\omega_{\rm{bath}}$.
Essentially, the singlet fidelity loss in this case can be understood by the Fermi Golden rule, where the MagR spins provide resonant perturbations that cause the singlet-triplet transition of the radical pair \cite{Tannoudji_1991}.
\crefformat{figure}{Figure~#2#1{(a)}#3}\cref{fig:figure3} shows an opposite example, in which the frequencies of the power spectrum peaks and the singlet-triplet transition are mismatched.
In this case, the radical pair spins can hardly transition from the singlet state to the triplet states. Thus, the singlet productivity is very close to unity and has a negligible geomagnetic direction dependence [\crefformat{figure}{Fig.~#2#1{(d)}#3}\cref{fig:figure3}].

Second, the energy splittings of the triplet states are crucial to the biocompass.
Assuming that the resonance condition mentioned above is satisfied, and the three triplet states are nearly degenerate ($\omega_{T_\alpha}\approx \omega_{T}$).
In this case, Eq.~\crefformat{equation}{#2(#1{})#3}\cref{eq:short-time} becomes
\begin{equation}
  P_S(t) \approx 1 - \gamma_e^2 \int_{-\infty}^\infty \left(\sum_{\alpha} \mathcal{S}_{\alpha}(\omega) \right) F(t,\omega,\omega_{S T}) d\omega,
\end{equation}
where $\omega_{S T}=\omega_S -\omega_T$ and the total power spectrum is
\begin{equation}
  \sum_{\alpha} \mathcal{S}_{\alpha}(\omega) = \frac{1}{2^N}\sum_{m,n} \vert \langle\psi_m\vert \delta \mathbf{b} \vert \psi_n \rangle \vert^2 \delta(\omega-\epsilon_{mn}).\label{eq:totalspectrum}
\end{equation}
Note that the total power spectrum depends on the magnitude of the field difference, which is insensitive to the geomagnetic field direction.
Although the eigenstates $\vert\psi_m\rangle$ and $\vert\psi_n\rangle$ in Eq.~\crefformat{equation}{#2(#1{})#3}\cref{eq:totalspectrum} depend on the geomagnetic field direction, this dependence could be rather weak, particularly when averaging over all eigenstates.
This result is verified by our numerical calculations [see \cite{Sup} regarding $\sum_{\alpha} \mathcal{S}_{\alpha}(\omega)$]. \crefformat{figure}{Figure~#2#1{(b)}#3}\cref{fig:figure3} shows the singlet fidelity of the radical pair when the three triplets are degenerate.
The MagR spins cause a remarkable transition from the singlet state to the triplet states. However, the sensitivity to the field direction is significantly reduced [see \crefformat{figure}{Fig.~#2#1{(d)}#3}\cref{fig:figure3}].
In sharp contrast, the nondegenerate case shows a strong magnetosensation ability [see \crefformat{figure}{Fig.~#2#1{(c)}#3}\cref{fig:figure3}].

\begin{figure}[b]
\includegraphics[scale=0.35]{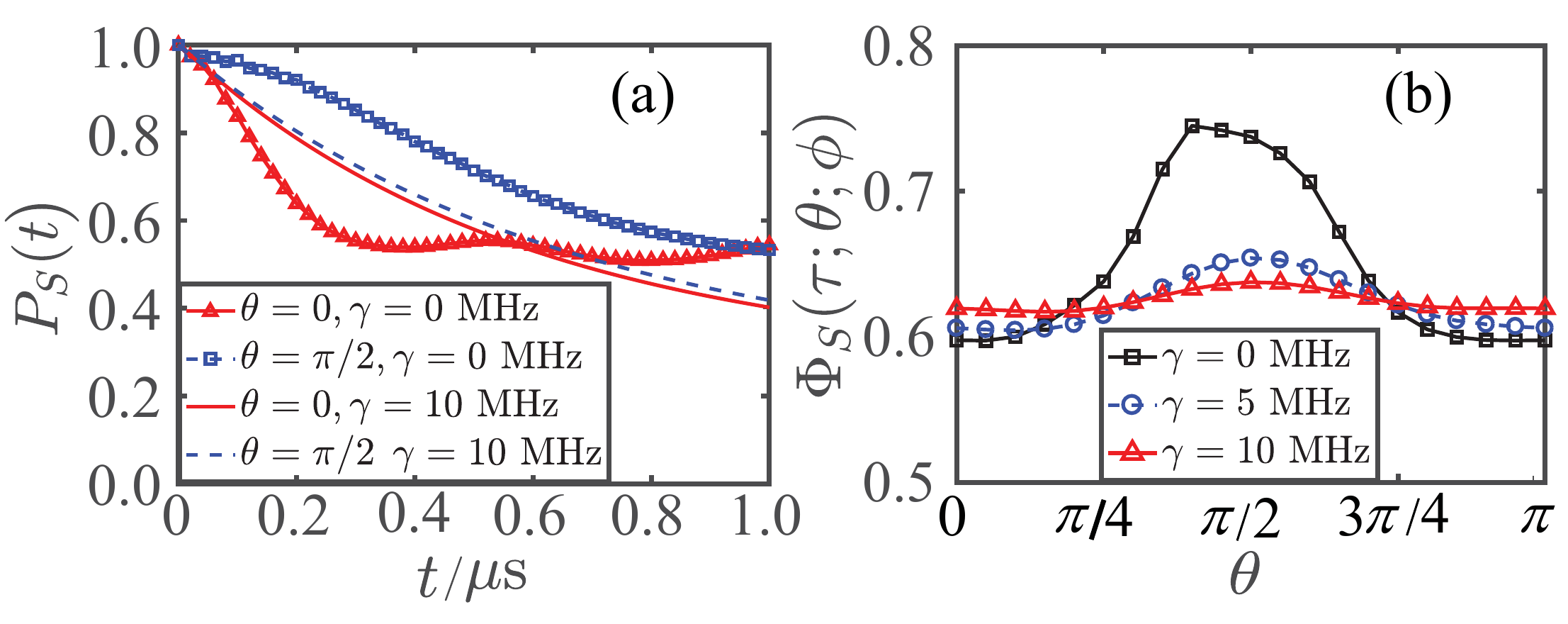}
\caption{\label{fig:figure4} (color online)
(a) The singlet fidelity for different relaxation and dephasing rates $\gamma$ for different geomagnetic field directions.
(b) The singlet productivity as a function of the geomagnetic field direction $\theta$ for different relaxation and dephasing rates.
}
\end{figure}

\emph{Incoherent effect.}---
Thus far, we have focused on the coherent dynamics of the radical pair spins and the MagR spins.
However, since the whole magnetosensation system is inevitably subjected to a biological environment (mainly via the electron-phonon interactions at the
room temperature), environment-induced decoherence must be considered.
Accordingly, we include the relaxation and decoherence of the MagR bath spins, which are governed by the Lindblad equation
\begin{equation}
  \dot{\rho}=  -i [H,\rho]+ \sum_{i=1}^{3N} \gamma_i\left\{ \hat{D}_i^\dagger\rho \hat{D}_i-\frac{1}{2} \hat{D}_i^\dagger \hat{D}_i \rho -\frac{1}{2} \rho \hat{D}_i^\dagger \hat{D}_i \right\}
  \label{eq:Lindblad},
\end{equation}
where $\hat{D_i} = \sigma_i^{(z)}$ and $\sigma_i^{(\pm)}$ (the Pauli matrices) for spin dephasing and spin relaxation processes, respectively, and $\gamma_i$ represents the corresponding relaxation and dephasing rates \cite{Lindblad_1976,Breuer_2002}.
For simplicity, we set $\gamma_i\equiv\gamma$ for all of the MagR spins.
\crefformat{figure}{Figure~#2#1{(a)}#3}\cref{fig:figure4} compares the singlet fidelity with and without the effect of environmental decoherence.
With the decoherence process described in Eq.~\crefformat{equation}{#2(#1{})#3}\cref{eq:Lindblad}, the radical pair can still undergo a singlet-triplet transition.
However, the geomagnetic field direction sensitivity is significantly reduced in an environment with strong decoherence, as shown in \crefformat{figure}{Fig.~#2#1{(b)}#3}\cref{fig:figure4}.
In this sense, the result indicates that the quantum coherence within the MagR is crucial to the biocompass.

\emph{Conclusion.}---
In summary, we establish a microscopic model of the magnetic-protein-assisted biocompass and analyze the physical origin of the magnetosensation.
With quantum mechanical calculations, we show that the magnetosensation of the radical pair is the consequence of the magnetic fluctuation of the MagR rather than the mean magnetization.
Furthermore, we discover that microscopic spin coupling and the level structure of the MagR and radical pair spins are essential to the magnetosensation.
We propose two general necessary conditions, a resonance condition and a nondegeneracy condition, for the biocompass.
These conditions provide more quantitative criteria for candidate biocompass systems and can be examined in future biophysical experiments at the molecular level with well-developed experimental electron spin resonance (ESR) and nuclear magnetic resonance (NMR) techniques.
We also find that quantum coherence plays an important role in the geomagnetic field navigation process.
This finding could inspire studies of various quantum effects in biological systems and bionic applications of artificial quantum systems.

\begin{acknowledgments}
We thank Prof. Chang-Pu Sun, Prof. Can Xie, Prof. Ren-Bao Liu and Dr. Yi-Nan Fang for their inspiring discussions. We  thank Prof. Hai-Guang Liu for sharing the knowledge of the MagR structure. We also thank Prof. Hai-Guang Liu, Prof. Jin Yu, Prof. Peng Zhang for their comments on the manuscript. The work was partially done while the third author worked at Peking University.
This work is supported by NSFC (Grant No. 11534002 and Grant No. 11671023) and NSAF (Grant No. U1930402 and Grant No. U1930403).
\end{acknowledgments}

% \bibliographystyle{apsrev4-1}
% \bibliography{MagR.bib}
\providecommand{\noopsort}[1]{}\providecommand{\singleletter}[1]{#1}%

\end{document}

\renewcommand{\theequation}{S\arabic{equation}}
\renewcommand{\thefigure}{S\arabic{figure}}
\renewcommand{\thetable}{S\arabic{table}}
\renewcommand{\bibnumfmt}[1]{[S#1]}
\renewcommand{\citenumfont}[1]{S#1}

\title{Supplemental Materials: Magnetic Noise Enabled Biocompass}
\author{Da-Wu Xiao}
\author{Wen-Hui Hu}
\affiliation{ Beijing Computational Science Research Center, Beijing 100193, China}%
\author{Yunfeng Cai}
\affiliation{ Cognitive Computing Lab, Baidu Research, Beijing 100085, China}%
\author{Nan Zhao}
\email{nzhao@csrc.ac.cn}
\affiliation{ Beijing Computational Science Research Center, Beijing 100193, China}%

\date{\today}

\maketitle
\renewcommand{\arraystretch}{1.5}
\begin{table}[htbp]

  \centering
  \fontsize{6.5}{8}\selectfont
  \caption{The coordinates of the MagR bath spins \cite{MagR1} and the radical pair spins. Here, $\mathbf{J}_i$ are the bath spins, and $\mathbf{S}_i$ are the radical pair spins. In our main text, we only consider the spins $\mathbf{J}_{5-12}$ for simplicity.}
  \label{tab:coordinates}
    \begin{tabular}{|c|c|c|c|c|c|c|c|}
    \hline
    \multirow{2}{*}{Spin}&
    \multicolumn{3}{c|}{Coordinates}&\multirow{2}{*}{Spin}&\multicolumn{3}{c|}{ Coordinates}
    \cr\cline{2-4}\cline{6-8}
    &x(\AA)&y(\AA)&z(\AA)&&x(\AA)&y(\AA)&z(\AA)\cr
    \hline\hline
    $\mathbf{J}_1$&11.4028& -3.63309& 83.1636&  $\mathbf{J}_2$ &10.7877& 4.58341& 77.0411\cr\hline
    $\mathbf{J}_3$&-10.8003& 3.79766& 82.7676&  $\mathbf{J}_4$ &-11.0965& -4.15934& 77.1673\cr\hline
    $\mathbf{J}_5$&-2.117&   11.5884& 28.8328&  $\mathbf{J}_6$ &-8.695& 7.23216& 24.3528\cr\hline
    $\mathbf{J}_7$&2.187& -11.5131& 29.0298&    $\mathbf{J}_8$ &8.9895& -7.46034& 23.9188\cr\hline
    $\mathbf{J}_9$&8.68875& 7.76766& -23.8662&  $\mathbf{J}_{10}$&1.9615& 11.3649& -29.3439\cr\hline
    $\mathbf{J}_{11}$&-8.9735&-7.76709&-23.8702&$\mathbf{J}_{12}$&-2.37825&-11.4266&-29.3517\cr\hline
    $\mathbf{J}_{13}$&-10.687&3.63041&-77.2724&$\mathbf{J}_{14}$&-11.286&-4.44509&-82.5264\cr\hline
    $\mathbf{J}_{15}$&11.2882&-3.62059&-77.4357&$\mathbf{J}_{16}$&10.728& 4.06066& -82.6077\cr\hline
    $\mathbf{S}_1$&20.8&12& 0&  $\mathbf{S}_2$&20.8&-12& 0\cr\hline
    \end{tabular}
\end{table}

\begin{figure}[htbp]
\includegraphics[scale=0.28]{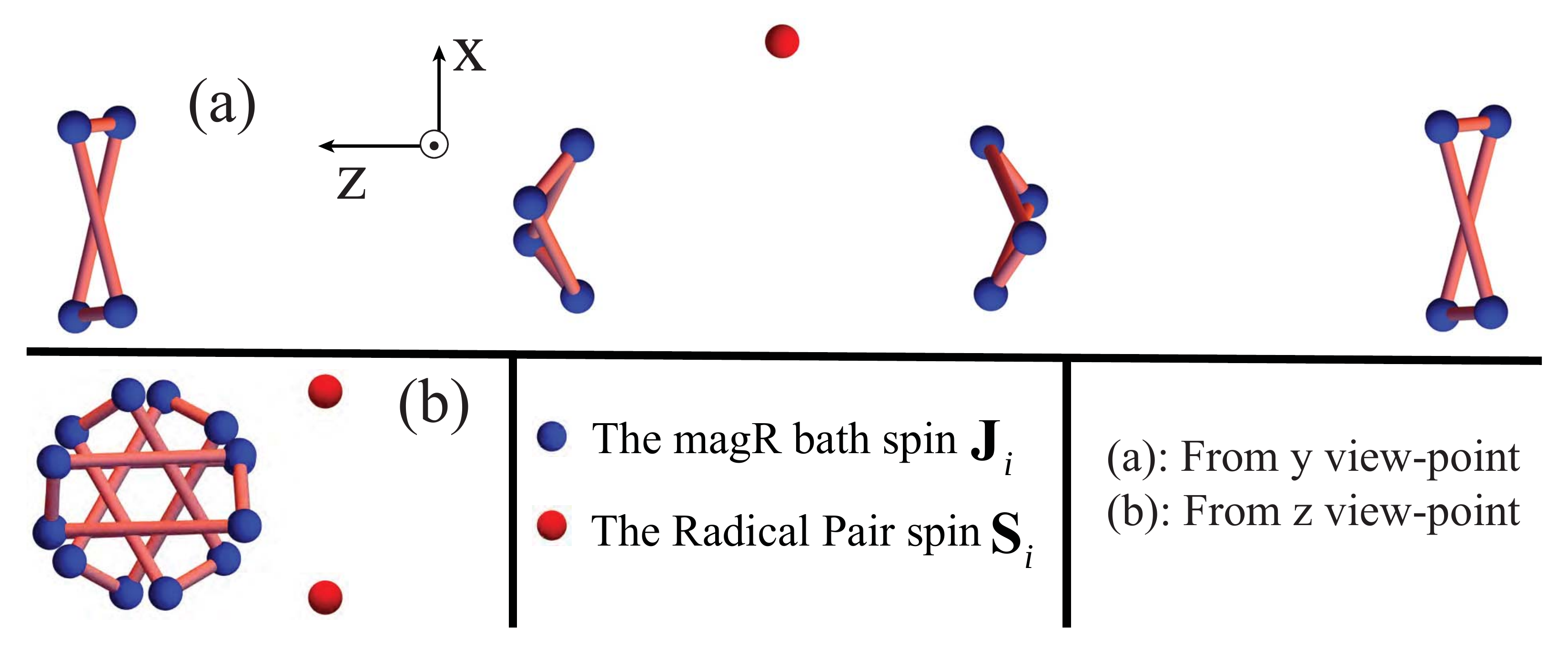}% Here is how to import EPS art
\caption{\label{fig:figure1} (color online).
Coordinate demonstration of the bath spins and the radical pair spins in Table (\ref{tab:coordinates}).
The MagR spin bath consists of several rings.
Within a ring, the distance between two spins is approximately $1\sim2~\rm{nm}$, while the distance between two spins is greater than $5~\rm{nm}$.
The average distance between the radical pair and the MagR spin bath is approximately $\sim10^1~\rm{nm}$, which is neither too far nor too close.
The distance between two radical pair spins is approximately $2.4~\rm{nm}$. We also study the effect of the radical pair spin positions in Fig. (\ref{fig:figure7}).
}
\end{figure}

\begin{figure}[htbp]
\includegraphics[scale=0.28]{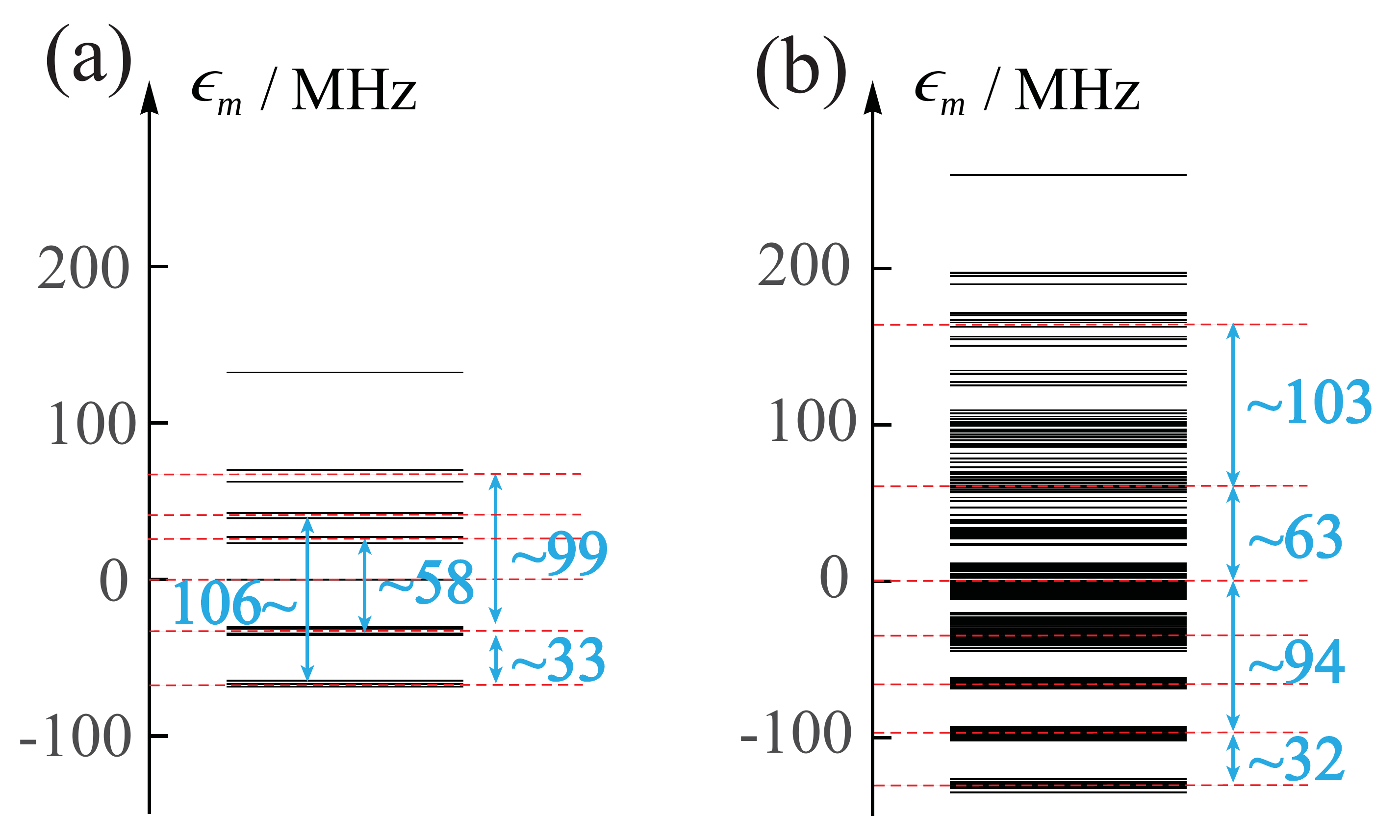}% Here is how to import EPS art
\caption{\label{fig:figure2} (color online).
(a) The energy spectrum of the MagR spin bath for one ring, i.e., $\mathbf{J}_{5-8}$. (b) The energy spectrum of the MagR spin bath for two rings, i.e., $\mathbf{J}_{5-12}$.
We show that the bath spin energy spectrum forms several bands, which correspond to the strong interaction within the rings.
The weak interaction between the rings splits the energy bands further.
Additionally, we notice that the characteristic splitting width of each energy band is larger than $>1~\rm{MHz}$, which sets the linewidth of the noise spectrum.
}
\end{figure}

\section{Short-time analysis of the singlet fidelity dynamics}
\begin{figure}[htbp]
\includegraphics[scale=0.45]{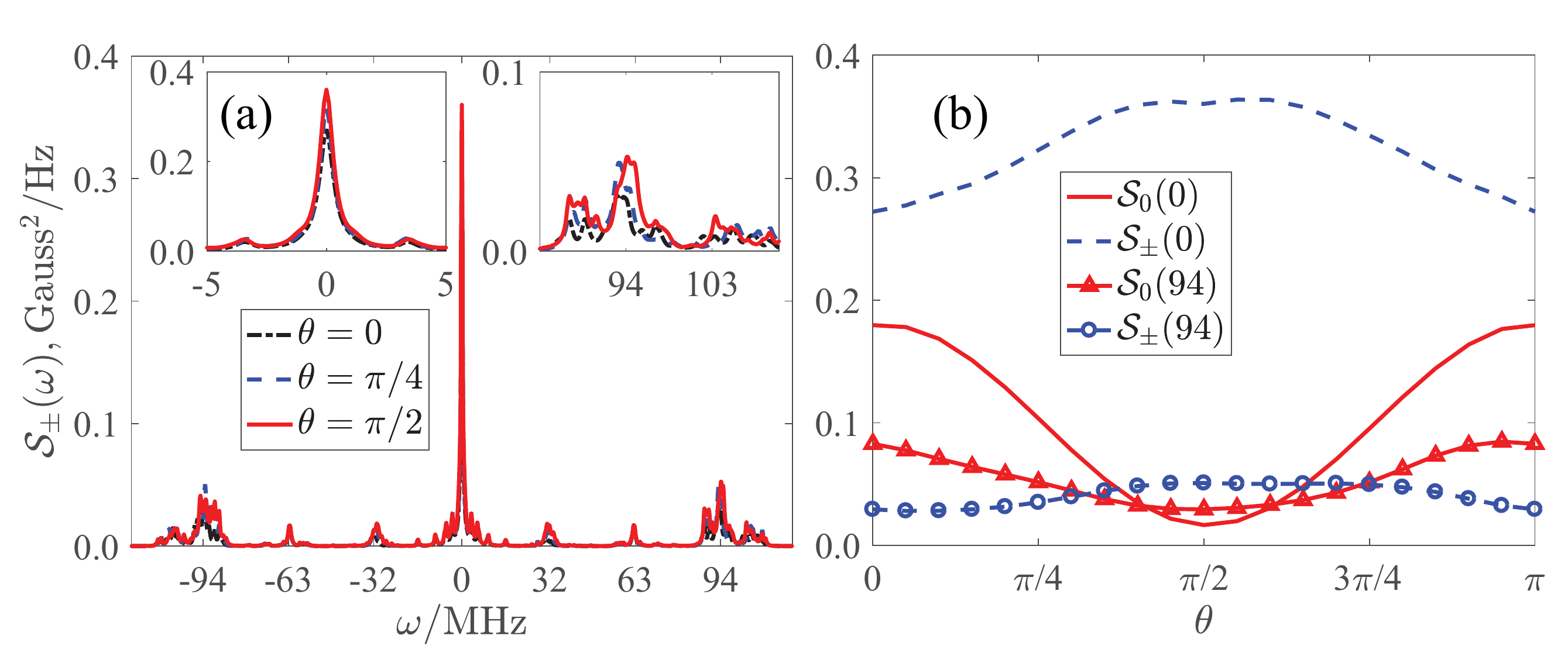}% Here is how to import EPS art
\caption{\label{fig:figure3} (color online).
(a) The noise spectrum component $\mathcal{S}_\pm(\omega)$ as a function of %Editor: Please ensure that the intended meaning has been maintained in this edit.
$\omega$. $\mathcal{S}_\pm(\omega)$ has similar behavior as $\mathcal{S}_0(\omega)$ shown in the main text; i.e., it has discrete peaks near $0,\pm32,\pm63,\pm64,$ and $\pm103~\rm{MHz}$.
The strength of the noise spectrum is also dependent on the geomagnetic direction.
(b) We show, when $\omega=0,94\rm{MHz}$, the noise spectrum $\mathcal{S}_{\pm,0}(\omega=0,94\rm{MHz})$ as a function of the geomagnetic direction $\theta$ to clarify the geomagnetic direction dependence of the noise spectrum.
The noise spectrum strength at the specified $\omega$ is very sensitive to the geomagnetic direction.
%Editor: Please ensure that the intended meaning has been maintained in this edit.
This signature is very important for a biocompass, which we have pointed out in the relevant discussion of our main text.
This result also sheds light on the mechanism of the radical pair biocompass model.
More attention should be paid to the noise spectrum for a realistic biological system, rather than the mean magnetization.
}
\end{figure}

\begin{figure}[htbp]
\includegraphics[scale=0.45]{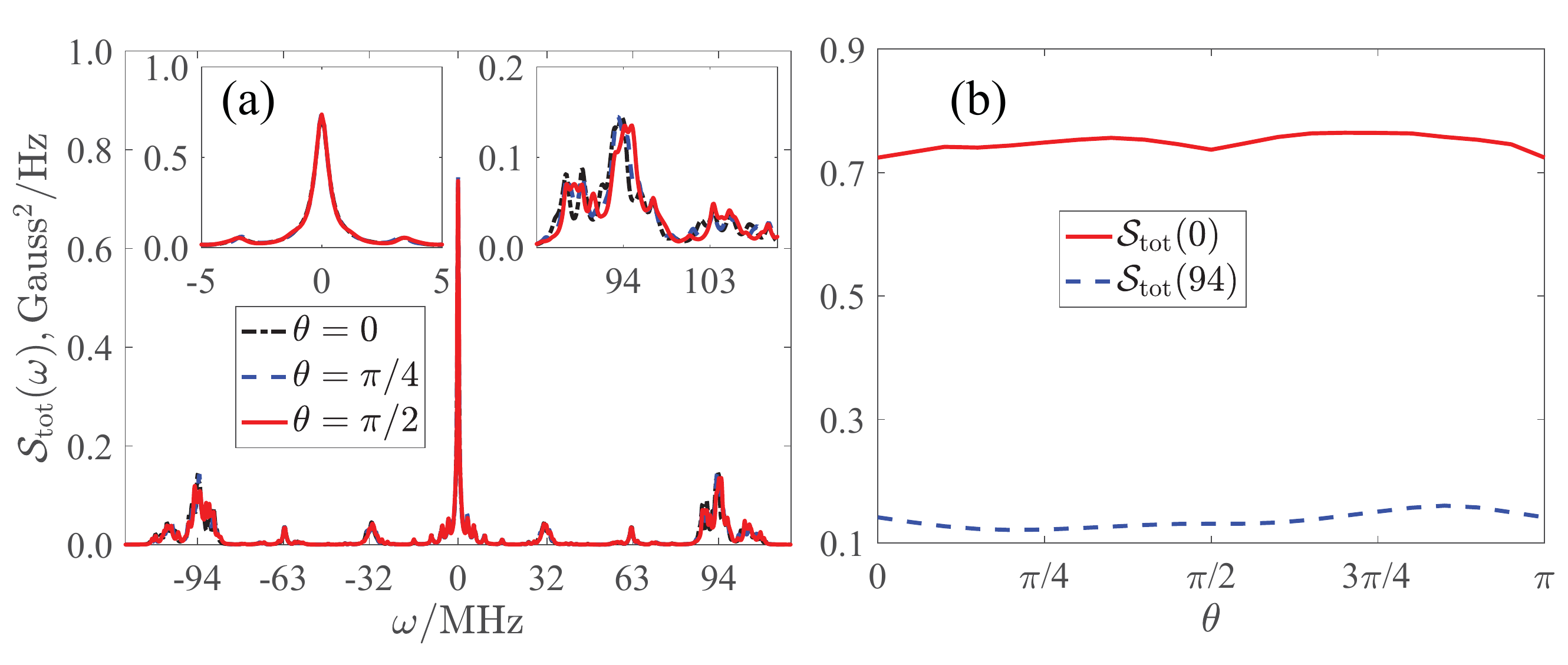}% Here is how to import EPS art
\caption{\label{fig:figure4} (color online).
In the main text, we have shown that the total noise spectrum $\mathcal{S}_{\rm{tot}}(\omega)\equiv\sum_{\alpha=0,\pm} \mathcal{S}_\alpha(\omega)$ is irrelevant with respect to the geomagnetic direction, which we demonstrate in Fig. (\ref{fig:figure4}).
(a) $\mathcal{S}_{\rm{tot}}(\omega)$ as a function of $\omega$ for different geomagnetic directions.
(b) $\mathcal{S}_{\rm{tot}}(\omega=0,94~\rm{MHz})$ as a function of the geomagnetic field direction $\theta$.
The geomagnetic direction dependence of $\mathcal{S}_{\rm{tot}}(\omega)$ is canceled.
However, one can still observe a very small dependence of the geomagnetic direction.
The geomagnetic field changes the Zeeman energy of $H_{\rm{bath}}$, which will consequently have a small effect on $\vert \psi_m \rangle$ and $\mathcal{S}_{\rm{tot}}(\omega)$.
}
\end{figure}

\begin{figure}[htbp]
\includegraphics[scale=0.45]{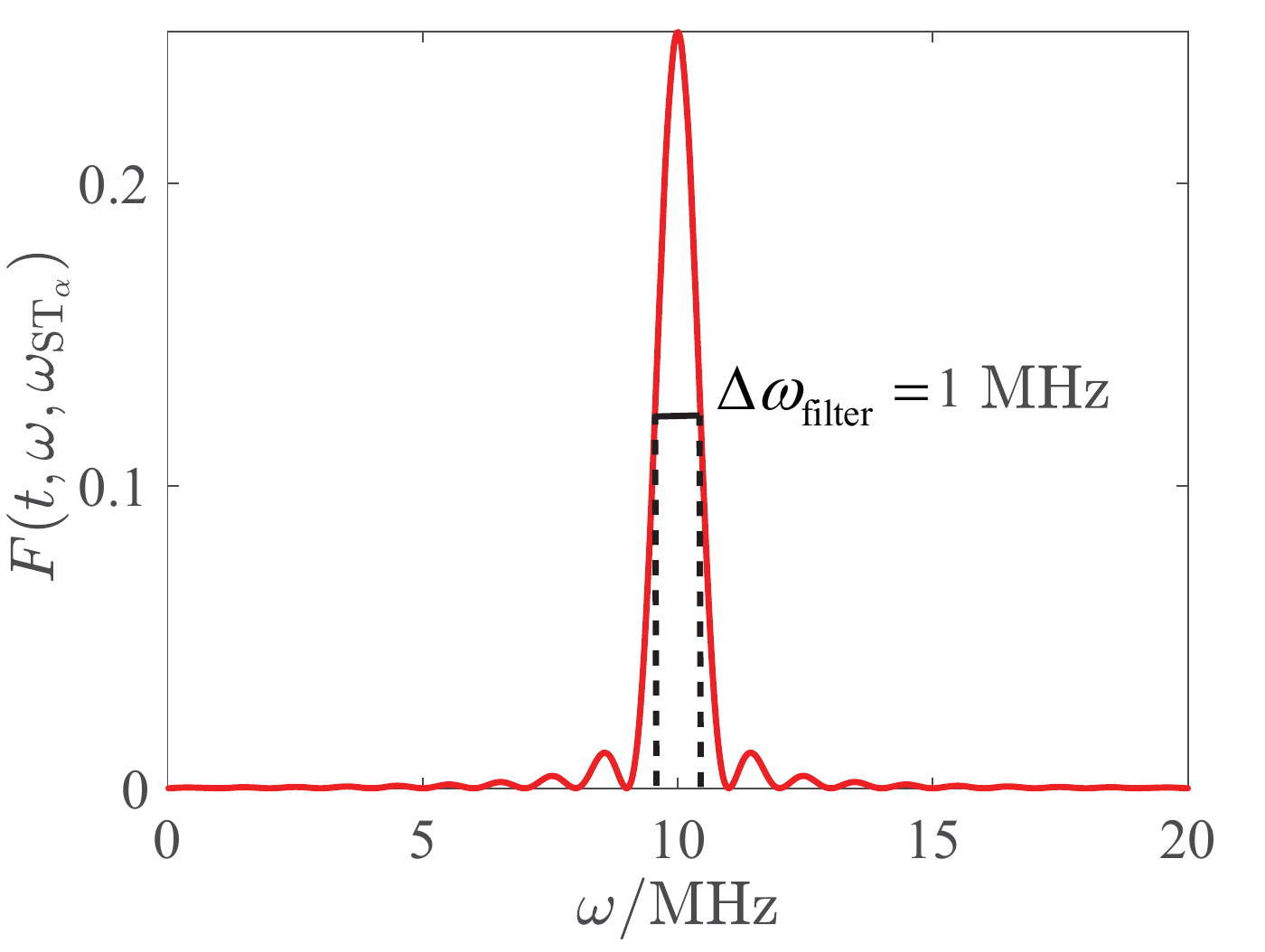}% Here is how to import EPS art
\caption{\label{fig:figure5} (color online).
The filter spectrum for $\omega_{\rm{S T_\alpha}}=-10~\rm{MHz}$ and $t=1~\mu s$ %Editor: Please ensure that the intended meaning has been maintained in this edit.
has sharp peaks near $\omega=-\omega_{\rm{S T_\alpha}}$, and the linewidth of the noise spectrum is approximately $1~\rm{MHz}$.
}
\end{figure}

\begin{figure}[htbp]
\includegraphics[scale=0.38]{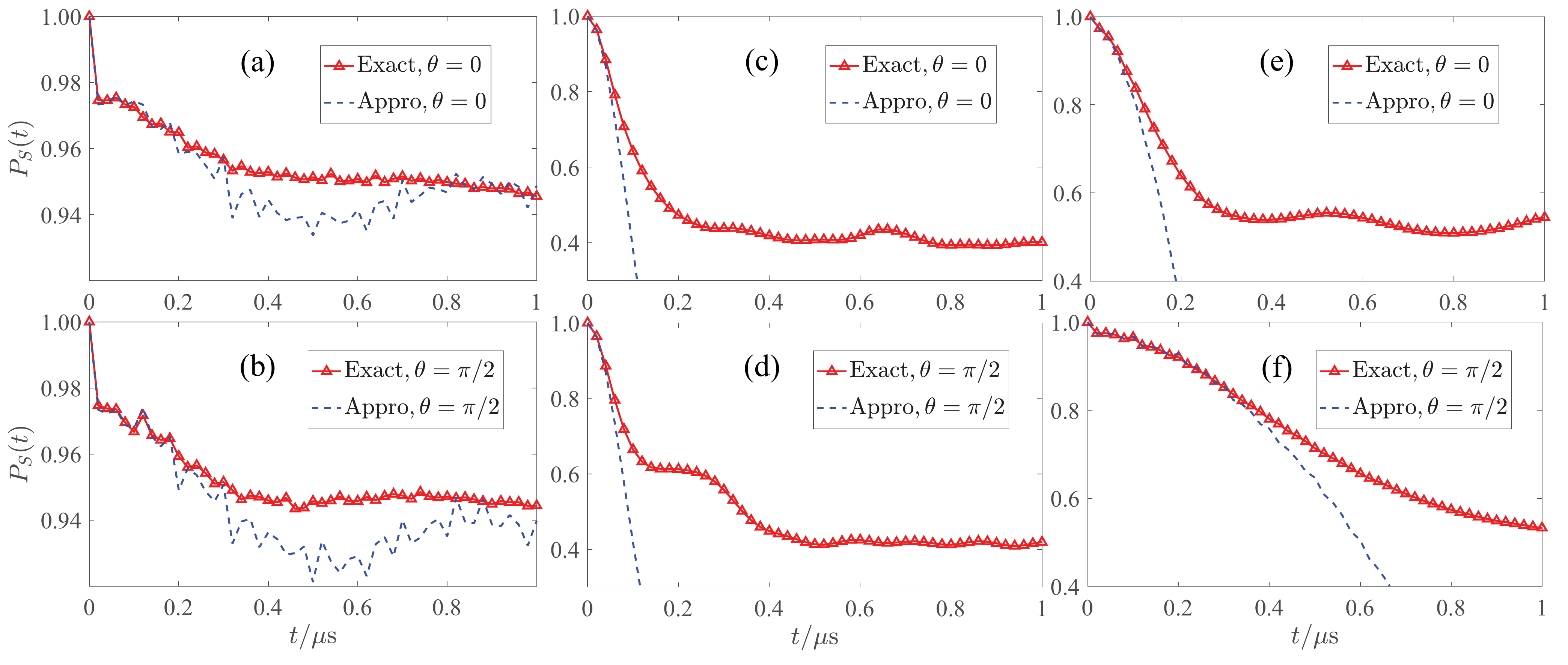}% Here is how to import EPS art
\caption{\label{fig:figure6} (color online).
We compare the approximation result and the exact numerical calculation of the singlet fidelity for all of the cases in Fig. (3) of our main text.
Here, (a,b), (c,d), and (e,f) correspond to the cases of Fig. (3a), Fig. (3b), % Editor: Add ", respectively" to the end of this sentence if direct correspondences are implied, e.g., if (a,b) corresponds to Fig. (3a).
and Fig. (3c) in our main text.
Our comparison shows that the short-time approximation fit very well with our exact numerical calculation.
}
\end{figure}

In this section, we provide a short-time analysis of the singlet fidelity dynamics \cite{shorttime,shorttimeZhao,Lindblad2}. As shown in the main text, we focus on the dynamics of the singlet fidelity (most symbols are defined in the main text)
\begin{align}
  P_S(t) &= {\rm Tr}\left[\vert S \rangle \langle S \vert~e^{-i H t}~\rho_{\rm{rp}}(0)\otimes\rho_{\rm{bath}}(0)~e^{i H t} \right] \nonumber \\
  & = \frac{1}{2^N}{\rm Tr}\left[\vert S \rangle \langle S \vert~e^{-i H t }~\vert S \rangle \langle S\vert~e^{i H t } \right],
\end{align}
where $1/{2^N}=\bigotimes_{i=1}^{N} 1/\rm{Tr}[{\mathbb I_i}]$ and ${\mathbb I_i}=1/2$ is the multiplicity of the bath spins. In the interaction picture with respect to $H_{\rm{rp}}+H_{\rm{bath}}$, the singlet fidelity can be approximated as
\begin{align}
  P_S(t)\approx1-\frac{1}{2^N} \int_0^t \int_0^{t_1} \rm{Tr}_{\rm{bath}}
  \left[ \langle S\vert V(t_2) V(t_1) + V(t_1)V(t_2) \vert S \rangle \right]dt_1 dt_2
  +\frac{1}{2^N}\int_{0}^{t}\int_{0}^{t}{\rm Tr}_{{\rm bath}}\left[\langle S\vert V\left(t_{2}\right)\vert S\rangle\langle S\vert V\left(t_{1}\right)\vert S\rangle\right],
\end{align}
where $V\left(t\right)$ is $H_{{\rm int}}$ in the interaction picture, defined by
\begin{equation}
  V\left(t\right)=\sum_{k=1,2}\gamma_{e}\mathbf{S}_{k}\left(t\right)\cdot\mathbf{b}_{k}\left(t\right)=\sum_{k=1,2}\gamma_{e}\left(e^{iH_{{\rm rp}}t}\mathbf{S}_{k}e^{-iH_{{\rm rp}}t}\right)\cdot\left(e^{iH_{{\rm bath}}t}\mathbf{b}_{k}e^{-iH_{{\rm bath}}t}\right).
\end{equation}
The singlet fidelity can be simplified as
\begin{align}
  P_S(t) &\approx 1-\frac{1}{2^N}\sum_{\alpha= {0,\pm}}\int_{0}^{t}\int_{0}^{t_{1}}{\rm Tr}_{{\rm bath}}\left[\left(\langle S\vert V\left(t_{2}\right)\vert T_{\alpha}\rangle\langle T_{\alpha}\vert V\left(t_{1}\right)\vert S\rangle+\langle S\vert V\left(t_{1}\right)\vert T_{\alpha}\rangle\langle T_{\alpha}\vert V\left(t_{2}\right)\vert S\rangle\right)\right]dt_{1}dt_{2} \nonumber \\
  &=1-\frac{1}{4}\frac{1}{2^N}\gamma_{e}^{2}\sum_{\alpha= {0,\pm}}\int_{0}^{t}\int_{0}^{t_{1}}{\rm Tr}_{{\rm bath}}\left[e^{i\omega_{{\rm S T_\alpha}}\left(t_{2}-t_{1}\right)}\left(\mathbf{n}_{\alpha}\cdot\delta\mathbf{b}\left(t_{2}\right)\right)\left(\mathbf{n}_{\alpha}\cdot\delta\mathbf{b}\left(t_{1}\right)\right)^{\dagger}+h.c.\right]dt_{1}dt_{2},
\end{align}
where $\omega_{{\rm S T_\alpha}}=\omega_S -\omega_{\rm{T_\alpha}},\mathbf{n}_{0}=\mathbf{n}_{\mathbf{B}}\equiv \mathbf{e}_z$, and $\mathbf{n}_{\pm}=(\mp\mathbf{e}_x- i \mathbf{e}_y)/\sqrt 2$.
In the eigenspace of $H_{\rm{bath}}$,
\begin{equation}
  H_{\rm{bath}}\vert \psi_m\rangle = \epsilon_m \vert \psi_m \rangle,
\end{equation}
the singlet fidelity can be further simplified as
\begin{align}
  P_S(t)&\approx1-\frac{1}{2}\gamma_{e}^{2}\sum_{\alpha={0,\pm}}\int_{0}^{t}\int_{0}^{t_{1}}\left(\sum_{m,n}\cos\left(\left(\epsilon_{\rm{mn}}+\omega_{{\rm S T_\alpha}}\right)\left(t_{2}-t_{1}\right)\right)\left|\langle \psi_m\vert\mathbf{n}_{\alpha}\cdot\delta\mathbf{b}(0)\vert \psi_n\rangle\right|^{2}\right)dt_{1}dt_{2} \nonumber \\
  & =1-\gamma_{e}^{2}\sum_{\alpha={0,\pm}}\sum_{m,n}\left|\langle \psi_m\vert\mathbf{n}_{\alpha}\cdot\delta\mathbf{b}(0)\vert \psi_n\rangle\right|^{2}\frac{\sin^2\left(\frac{\epsilon_{\rm{mn}}t+\omega_{{\rm S T_\alpha}}t}{2}\right)}{(\epsilon_{\rm{mn}}+\omega_{\rm{S T_\alpha}})^{2}},
\end{align}
where $\epsilon_{\rm{mn}}=\epsilon_m - \epsilon_n$. Define the noise spectrum as
\begin{equation}
  \mathcal{S}_\alpha(\omega)=\frac{1}{2^N}\sum_{m,n}\vert \langle\psi_m\vert \mathbf{n}_\alpha\cdot \mathbf{\delta b}(0) \vert \psi_n \rangle \vert^2 \delta(\omega-\epsilon_{\rm{mn}}),
\end{equation}
and we have
\begin{equation}
  P_S(t) =1 - \sum_{\alpha={0,\pm}} \gamma_e^2 \int_{-\infty}^\infty \mathcal{S}_\alpha(\omega)  F(t,\omega,\omega_{\rm{S T_\alpha}}) d\omega
\end{equation}
with
\begin{equation}
F(t,\omega,\omega_{\rm{S T_\alpha}})=\frac{\sin^2\left(\frac{\omega t+\omega_{{\rm S T_\alpha}}t}{2}\right)}{(\omega+\omega_{\rm{S T_\alpha}})^{2}}.
\end{equation}
Fig. (\ref{fig:figure3}), Fig. (\ref{fig:figure4}) and Fig. (2c) in our main text study the properties of the noise spectrum in detail.
In Fig. (\ref{fig:figure5}), we study the properties of the filter spectrum $F(t,\omega,\omega_{\rm{S T_\alpha}})$ in detail.
We also compare the approximation result with the exact numerical result of the singlet fidelity in Fig. (\ref{fig:figure6}), which shows that the approximation works very well on a short time scale.

\begin{figure}[htbp]
\includegraphics[scale=0.38]{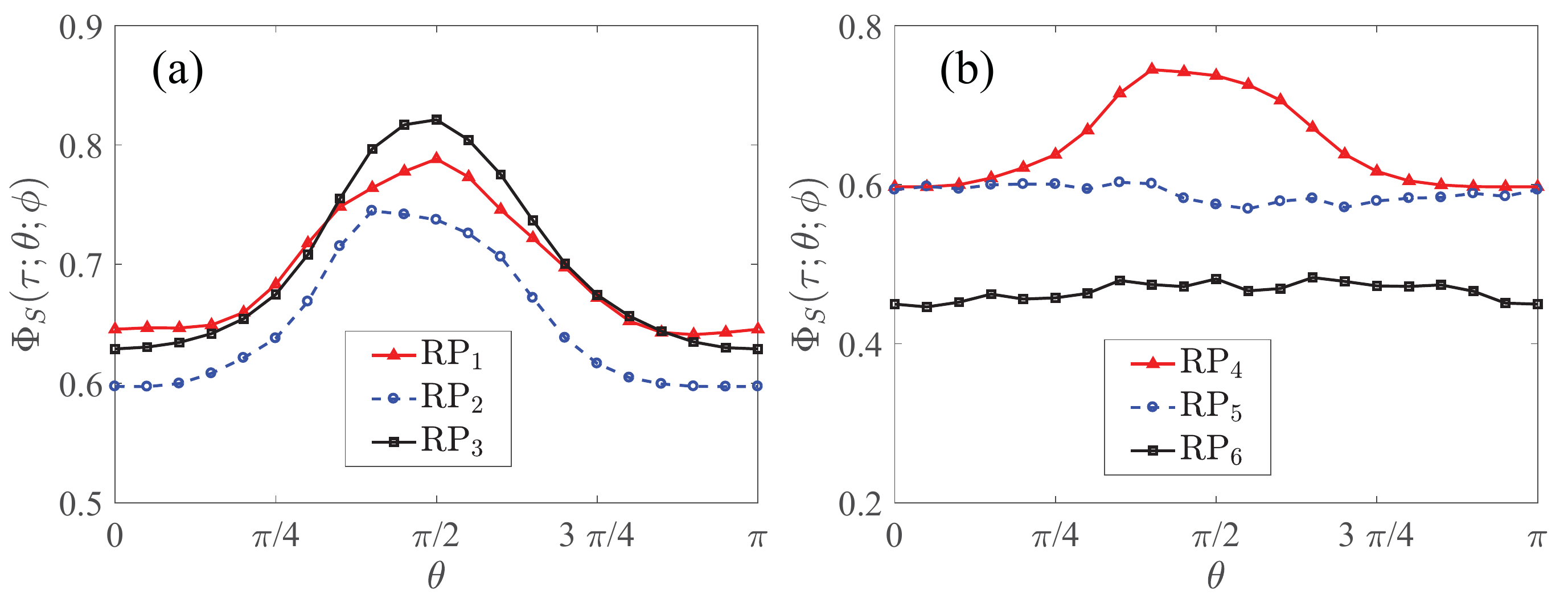}% Here is how to import EPS art
\caption{\label{fig:figure7} (color online).
In our model, the position of the radical pair electron spins is arbitrarily chosen because of the lack of information about the radical pair spin positions.
Here, we study the effect of the radical pair spin positions on the biocompass sensitivity (which is the singlet productivity in our model).
(a) The singlet productivity behavior when the distance between two radical pair spins changes, where $\rm{RP}_1,\rm{RP}_2,\rm{RP}_3$ represents the position of the radical pair spins $\mathbf{r}_{\mathbb{S}_1}=(20.8,6,0),(20.8,12,0),(20.8,18,0)$ and $\mathbf{r}_{\mathbb{S}_2}=(20.8,-6,0),(20.8,-12,0),(20.8,-18,0)$
in units of $\AA$.
(b) The singlet productivity behavior when the orientation (with the same distance) between two radical pair spins changes, where $\rm{RP}_4,\rm{RP}_5,\rm{RP}_6$ represents the position of the radical pair spins $\mathbf{r}_{\mathbb{S}_1}=(20.8,12,0),(20.8,8.48,8.48),(20.8,0,12)$ and $\mathbf{r}_{\mathbb{S}_2}=(20.8,-12,0),(20.8,-8.48,-8.48),(20.8,0,-12)$
in units of $\AA$.
Our results show that the behavior of $\Phi_S(\tau;\theta;\phi)$ is very sensitive to the orientation of the radical pair, while the distance between the two radical pair spins only slightly adjusts the biocompass sensitivity.
Our results also suggest that detailed information about the radical pair spin positions is very important for a biocompass model.
}
\end{figure}

\begin{figure}[htbp]
\includegraphics[scale=0.38]{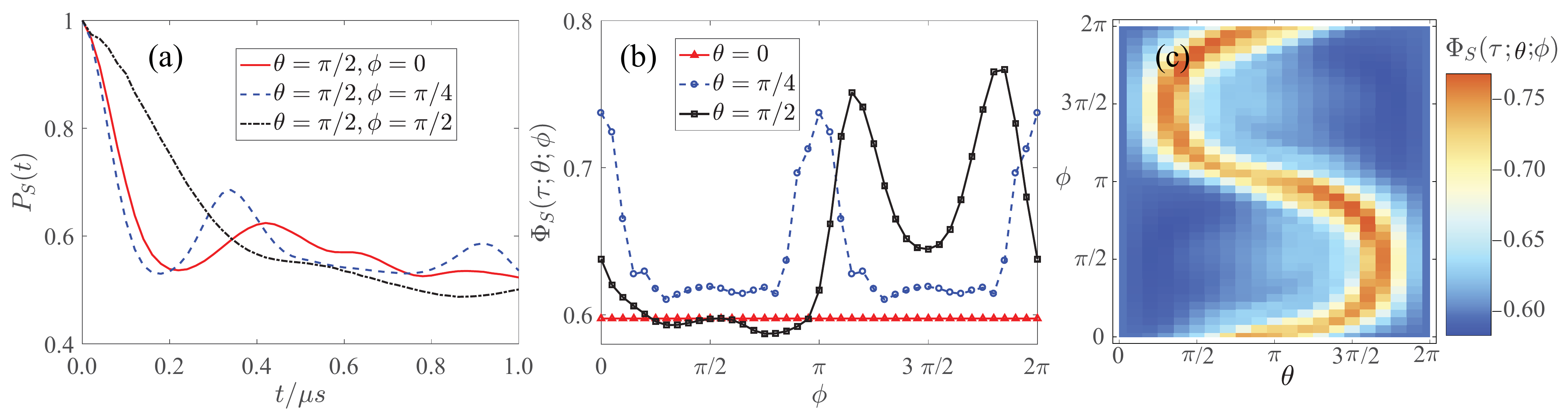}% Here is how to import EPS art
\caption{\label{fig:figure8} (color online).
In the main text, we only consider the singlet fidelity as a function of the axial angle of the geomagnetic field, $\theta$.
Now, we study the dynamic sensitivity and biocompass sensitivity as a function of the azimuthal angle of the geomagnetic field, $\phi$. Here, the energy structures of the radical pair spins are the same as those in Fig. (3c).
(a) The singlet fidelity behavior for different $\phi$.
(b) The singlet productivity as a function of $\phi$ for different $\theta$.
(c) The singlet productivity density plot as a function of $\theta$ and $\phi$.
Our result shows that the biocompass also works for sensing the azimuthal angles $\phi$.
These results are very important to the radical pair model in terms of explaining the navigation mechanism of immigrating animals, because sensitivity %Editor: Please ensure that the intended meaning has been maintained in this edit.
% Editor: This phrase is not clear, and a revision is necessary. Consider "distinguish the back and forth aspects of immigration" or "distinguish back and forth aspects from immigration".
to $\phi$ means that animals can distinguish back and forth in immigration.
}
\end{figure}

%% To AJE's Editor: Don't need to edit after this line.
\newpage
\section{The Effect of the Nuclear Spin}

Generally speaking, the nuclear spins provide a fluctuating magnetic field via the hyperfine couplings to the radical pair. Since the nuclear spins are unpolarized, the mean value of the fluctuating field is zero, but the variance of the field is finite. In this sense, the effect of the nuclei can be understood by the similar picture of MagR electron spins as we studied in the manuscript, i.e., \textit{the magnetic noise enabled singlet-triplet transition}. Nevertheless, the nuclear spins differ from the electron spins by their $\sim10^{-3}$ times smaller magnetic moment, their hyperfine couplings to the radical pair and their spatial distribution. Indeed, the nuclear spin effect was studied both theoretically and experimentally [Refs. \cite{Cai,Solov,Maeda}]. Here, we provide an analysis on the nuclear spin effect on the biocompass, with a comparison with the electron spin bath of MagR.

In principle, a radical pair coupled to a nuclear spin bath can be described by the following Hamiltonians:
\begin{equation}
H = H_{\rm rp} + H_{\rm bath} + H_{\rm hf}.
\end{equation}
The Hamiltonian of the radical pair $H_{\rm rp}$ is the same as Eq. (2) in the main text. The spin bath Hamiltonian is
\begin{equation}
H_{\rm bath} =  \sum_{i=1}^{N} \gamma_i ~ \mathbf{B} \cdot \mathbf{I}_i + \sum_{i>j=1}^{N} \mathbf{I}_i \cdot \mathbb{D}_{ij} \cdot \mathbf{I}_j,
\end{equation}
where $\{\mathbf{I}_i\}_{i=1}^N$ are nuclear spins, $\mathbb{D}_{ij} $ is the dipolar coupling tensor. The hyperfine coupling is described by
\begin{equation}
H_{\rm hf}=\sum_{k, i}^{N} \mathbf{S}_k \cdot \mathbb{A}_{k i}\cdot \mathbf{I}_{i} \equiv \sum_{k} \mathbf{S}_k \cdot \mathbf{b}_k^{(n)},
\label{eq:hf}
\end{equation}
where $\mathbb{A}_{k i}$ is the hyperfine tensor, $\mathbf{b}_k^{(n)}$ is the nuclei-induced magnetic field to $\mathbf{S}_k$. The hyperfine interaction $H_{\rm hf}$ consists of two parts: the Fermi contact interaction
\begin{equation}
  H_{F}=\sum_{k, i}^{N}A_{k i}^{(F)} \mathbf{S}_{k} \cdot \mathbf{I}_{i}\equiv\sum_{k}\mathbf{b}_k^{(F)}\cdot\mathbf{S}_k
\end{equation}
and the dipole-dipole coupling
\begin{equation}
  H_{D}=\sum_{k, i}^{N}\mathbf{S}_{k} \cdot \mathbb{A}_{k i}^{(D)} \cdot \mathbf{I}_{i}\equiv\sum_{k}\mathbf{b}_k^{(D)}\cdot\mathbf{S}_k.
\end{equation}
Here, we define the Fermi contant interaction induced magnetic field $\mathbf{b}_k^{(F)}$ and the dipole-dipole interaction induced magnetic field $\mathbf{b}_k^{(D)}$.
Note that the Fermi contact interaction is isotropic, while the dipole-dipole interaction is anisotropic.

\begin{figure}
\includegraphics[scale=0.4]{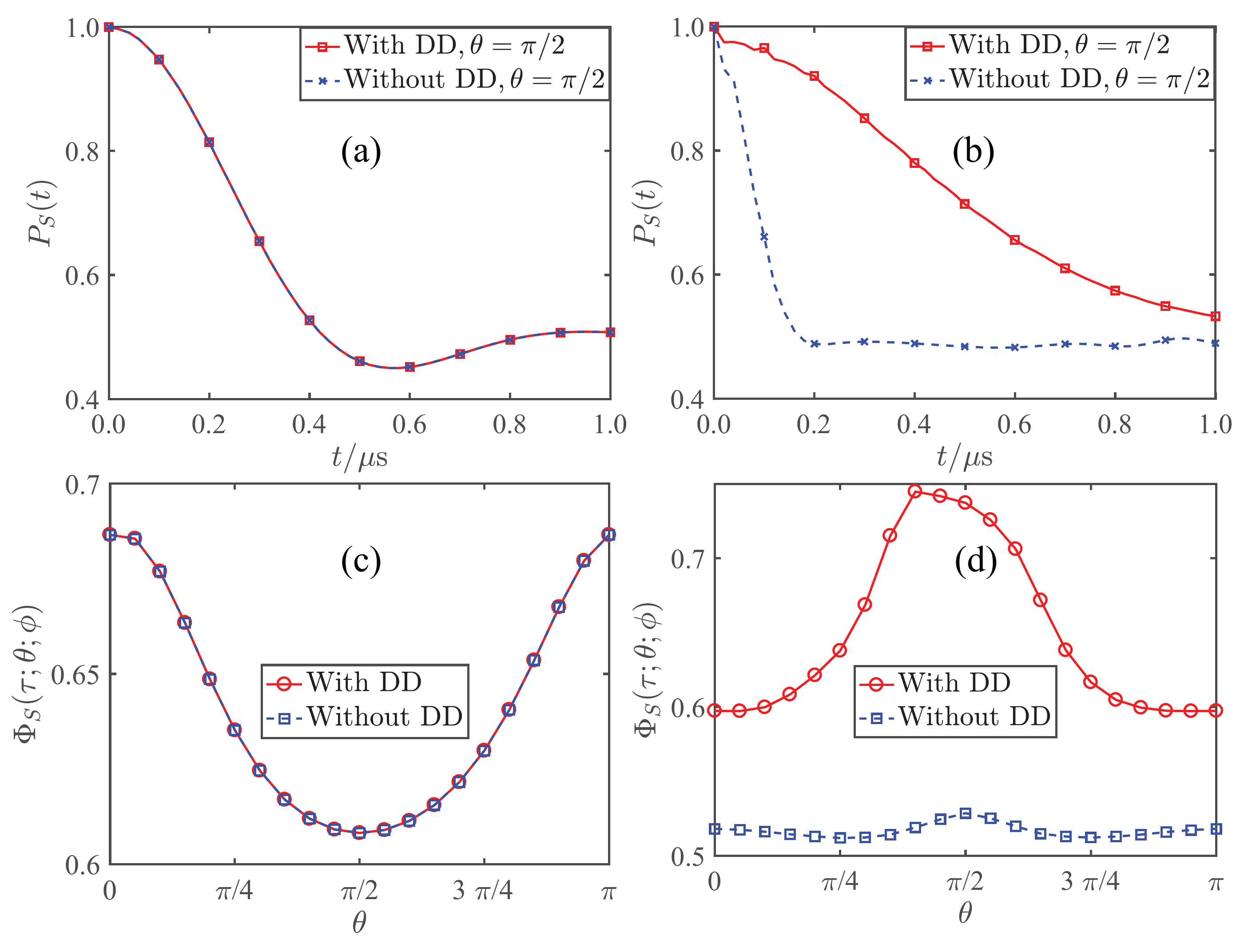}
\caption{(color online).
(a) The dynamics of the singlet fidelity $P_S(t)$ between the nuclear spin bath with/without the dipole-dipole interaction within the spin bath. The dipole-dipole interaction within the nuclear spin bath has negelible effect on $P_S(t)$ in $1\,\rm \mu s$.
(b) The singlet productivity $\Phi(\tau,\theta,\phi)$ as a function of the geomagnetic field direction $\theta$ with $\phi=0$. The dipole-dipole interaction within the nuclear spin bath is unimportant for the biocompass.
(c) The same as (a) but for the electron spins bath. The dipole-dipole interaction is strongly affects the dynamics of the singlet fidelity.
(d) The same as (c) but for the electron spins bath. The dipole-dipole interaction is crutial for the biocompass.\label{fig:figure10}}
\end{figure}

Although the Hamiltonians above have similar forms to the Hamiltonians of an electron spin bath [i.e., Eqs. (1)-(4) in the main text], the relative strength of $H_{\rm hf}$ [or the Hamiltoinan $H_{\rm int}$ for the electron spin bath, see Eq.(4) in the main text] and $H_{\rm bath}$ is quite different.
To characterize this difference, we define two ratios $\eta_{\rm n}=\frac{||H_{\rm hf}||}{|| H_{\rm bath} ||}$ and $\eta_{\rm e}=\frac{||H_{\rm int}||}{|| H_{\rm bath} ||}$, where $||\cdot||$ denotes the matrix norm.
\begin{itemize}
\item For nuclear spin bath, the interactions within the bath Hamiltonian $H_{\rm bath}$ is much weaker than the hyperfine coupling $H_{\rm hf}$ to the radical pair, i.e., $\eta_n\gg1$.

In the bath Hamiltonian $H_{\rm bath}$, the Zeeman energy of a nuclear spin in geomagnetic field is in the order of $\sim\,\rm{kHz}$, and the typical dipole-dipole interaction between two nuclear spins (e.g., two hydrogen nuclear spins separated by $2\,\AA$) is $\sim 10^1\,\rm{kHz}$ at the most. However, the Fermi contact coupling in $H_F$ could reach $\sim 10^3\,\rm{kHz}$ or even stronger (depending on the electron spin density at the nuclei), and the dipole-dipole interaction in $H_D$ could be $\sim 10^2\,\rm{kHz}$ for neighboring nuclei.

\item 	For electron spin bath, the interaction within the bath Hamiltonian $H_{\rm bath}$ is typically stronger than the coupling $H_{\rm int}$ to the radical pair, i.e., $\eta_e\ll1$.

Note that both interactions within $H_{\rm bath}$ and $H_{\rm int}$ are dipole-dipole couplings, which depends on the distance $r_{ij}$ between two electron spins $\mathbf{J}_i$ and $\mathbf{J}_j$ as $r_{ij}^{-3}$. The distance between spins within the MagR ($\sim$ 2 nm for a Fe-S ring) is much smaller than the distance between the MagR spins and the radical pair spins ($\sim$ 10 nm in our model). Consequently, the interaction within $H_{\rm bath}$ can reach $\sim\,10^3\,\rm{kHz}$, while the strength of $H_{\rm int}$ is in the order of $\sim 10^2\,\rm{kHz}$.
\end{itemize}
With the analysis of the coupling strength ratios, an obvious difference between nuclear spin bath and electron spin bath is the importance of interactions within the bath. Figure (\ref{fig:figure10} (a) \& (c)) shows the numerical results of the evolution of the singlet population with and without the nuclear spin dipolar interaction. In this case, the dipole-dipole coupling between the bath spins ($< 10\,\rm{kHz}$) is too week to produce any observable effect in the relevant time scale ($\sim 1\,\mu s$), which agrees with the previous studies [Refs. \cite{Cai2010,Hore,Nature}]. While, in the case of electron spin bath, the dipole-dipole coupling between bath spins strongly affect the dynamics of the singlet population and total singlet productivity [see Fig. \ref{fig:figure10}(b) \& (d)].

Because of the different relative strength between the nuclear spin bath and the electron spin bath, we need different theoretical tools..
The small ratio $\eta_e\ll1$ for the electron spin bath allows the perturbation treatment of the interaction $H_{\rm int}$. As we have shown in our manuscript, the singlet-triplet transition caused by MagR can be well understood by the picture of noise spectrum and the Fermi golden rule, which is essentially a perturbation treatment. However, the fact $\eta_n\gg1$ for the nuclear spin bath indicates the dynamics of the nuclear spins are strongly affected by their coupling to the radical pair. A simple perturbative picture in this case is usually not available, and a full quantum mechanical calculation has to be performed as shown in Refs. \cite{Cai2010,Cai,Solov,Maeda}.

Indeed, the sharp contrast between electron spin bath and nuclear spin bath was studied in the system of solid-state spin qubit. Experimental [Ref. \cite{Science}] and theoretical [Ref. \cite{PRB}] works have shown that the effect of electron spin bath can be well modeled by a stochastic process. While, in the nuclear spin bath case, the back-action of the hyperfine coupling causes counter-intuitive behavior of the central spin [Refs. \cite{Nan,NC}]. Here, the biocompass system and the radical pair model provides a second example to elucidate the different environmental effect between electron and nuclear spin baths.

Besides the difference of $H_{\rm bath}$ between the nuclear spin bath and the electron spin bath, the interaction $H_{\rm hf}$ in the nuclear spin bath is also crucial to the magnetic biocompass. As we have shown in Eq. (\ref{eq:hf}), in additional to the dipole-dipole coupling $H_D$, the hyperfine interaction $H_{\rm hf}$ between the nuclear spin bath and the radical pair contains the Fermi contact coupling $H_F$, which is not present in the electron spin bath case. To study the role of $H_D$ and $H_F$ of the nuclear spin bath, we first consider a toy model with a single nuclear spin  with the following Hamiltonian:
\begin{equation}
  H_{\rm toy}=\gamma \mathbf{B}\cdot(\mathbf{S}_1+\mathbf{S}_2)+(A_F\, \mathbf{S}_1 + \mathbf{S}_1\cdot\mathbb{D})\cdot\mathbf{I},
\end{equation}
where $A_F$ is the Fermi contact interaction strength and $\mathbb{D}=\frac{\mu_0\gamma_e\gamma_n\hbar(1-3\hat{\mathbf{r}}\hat{\mathbf{r}})}{4\pi r^3}$ is the dipole-dipole coupling tensor. We ignored the finite size of the electron wavefunction, and use the point dipole approximation to model the anisotropic coupling.  For neighboring nuclei, the Fermi contact interaction could be much stronger than the dipole-dipole coupling. However, note that the Fermi contact coupling is of isotropic nature. With the Fermi contact coupling only, the nuclear spin bath cannot provide the magnetosensation of the direction of geomagnetic field.

\begin{figure}
\includegraphics[scale=0.4]{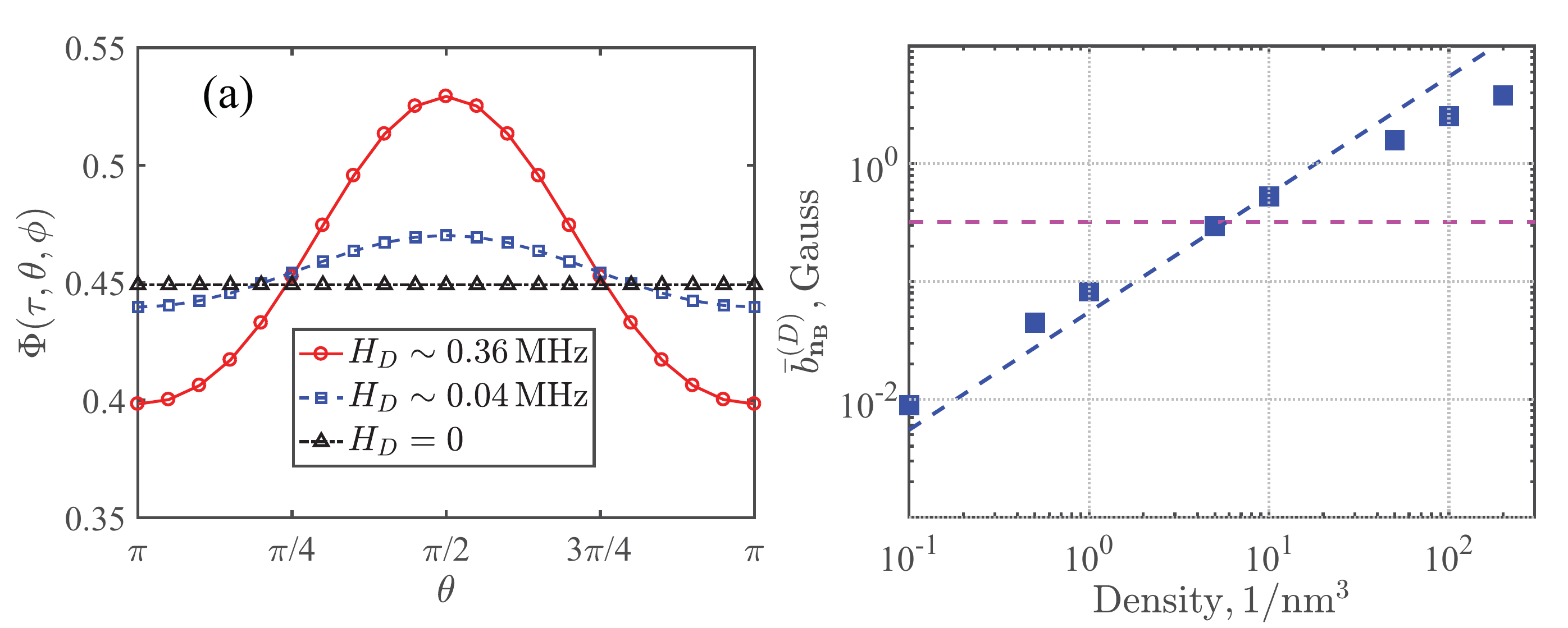}
\caption{(Color Online)  (a) The singlet fidelity $\Phi(\tau,\theta,\phi)$ for different different dipolar coupling strength of $H_D$. The Fermi contact coupling strength is $2.8\,\rm{MHz}$.
(b) The average dipole field $\mathbf{b}_{k}^{(D)}$ along the geomagnetic field direction as a function of the concentration of the nuclear spin bath.  The horizontal line indicates the typical fluctuating field strength provided by the MagR electron spins as discussed in the main text.\label{fig:figure11}}
\end{figure}

Figure (\ref{fig:figure11}a) shows the singlet productivity as a function of the magnetic field direction for a given Fermi contact strength $A_F=2.8\,\rm{MHz}$ and various dipole-dipole coupling strength. In the absence of dipole coupling $H_D=0$, the singlet productivity is \textit{independent} of the field direction, and increasing the dipole coupling strength enhances the magnetosenstation ability. With this toy model, we find that, despite of the strong Fermi contact coupling, the dipole-dipole coupling plays a key role in the biocompass process. For the radical pair in a nuclear spin bath, the magnetosensation of the geomagnetic field originates from the anisotropicity of the spin dipole-dipole interaciton.

To have a more quantitative understanding of the dipole-dipole interaction strength of the nuclear spins, we calculate the projection of the dipole field $\mathbf{b}_k^{(D)}$ (with the dipole-approximation $\mathbb{D}_{ki}=\frac{\mu_0\gamma_e\gamma_n\hbar(1-3\hat{\mathbf{r}}_{ki} \hat{\mathbf{r}}_{ki})}{4\pi r_{ki}^3}$
) along the geomagnetic field direction, and average the results over different random bath configurations of a given spin concentration (the number of nuclear spins per unit volume). Figure (\ref{fig:figure11}b) shows that the averaged dipole field strength is proportional to the bath spin concentration. This is because the dipole field strength and the bath spin concentration both scale with the inter-spin distance $r$ as $r^3$.

Comparing the strength of the dipole field produced by the MagR electron spins and the dipole field strength of nuclear spins, we find that, with a concentration of the order $\sim 10^1\,\rm{nm}^3$, the nuclear spin bath provides a dipole field with comparable strength to the MagR electron spins ($\sim 10^{-1} - 10^{0} $ Gauss ). The concentration of $\sim 10^1\,\rm{nm}^3$ is close to the possible nuclear spin concentration in biology systems, which implies that both MagR electron spin bath and the nuclear spin bath around the radical pair can have comparable contribution to the biocompass process.

In summary, we investigate the role of nuclear spin bath by the full quantum mechanical calculations and model analysis. Particularly, we clarify the key difference between the nuclear spin bath and the electron spin in the biocompass process. We show that, for nuclear spin bath, the hyperfine coupling $H_{\rm hf}$ dominates the quantum dynamics; While, for the electron spin bath, the spin interaction within the bath (i.e., the bath Hamiltonian $H_{\rm bath}$) is crucial to understand noise enabled biocompass behavior. Furthermore, we analyze the different role of the isotropic Fermi contact coupling $H_F$ and the anisotropic dipolar coupling $H_D$ of the nuclear spin bath. We show that the nuclear spin bath could provide a comparable dipole field strength to the MagR electron spin bath.